\documentclass[a4paper,8pt]{article}
\pdfoutput=1 
\usepackage{jheppub_mod}

\usepackage{ulem}
\usepackage{afterpage}

\usepackage{wrapfig}
\graphicspath{{Figs_new/}}
\usepackage{graphicx}
\usepackage{subfig}
\usepackage{appendix}

\def\mo{\ifmmode^{-1}\else$^{-1}$\fi}
\def\,{\thinspace}
\def\MJysr{\ifmmode \,$MJy\,sr\mo$\else \,MJy\,sr\mo\fi}
\def\Mpc{\ifmmode $\,Mpc\mo$\else\,Mpc\mo\fi}
\def\Mpch{\ifmmode $\,Mpc\,h\mo$\else\,Mpc\,h\mo\fi}

\def\deg{\ifmmode^\circ\else$^\circ$\fi}
\def\Msolar{\ifmmode{\rm M}_{\mathord\odot}\else${\rm M}_{\mathord\odot}$\fi}
\def\Mstar{\ifmmode{\rm M}_{\star}\else${\rm M}_{\star}$\fi}

\newcommand{\OVI}{\ensuremath{O\mathtt{VI}}}
\newcommand{\planck}{{\it Planck\/}}
\newcommand{\Nside}{\ensuremath{N_{\rm side}}}

\newcommand{\fsky}{\ensuremath{f_{\rm sky}}}
\newcommand{\Npairs}{\ensuremath{N_{\rm pairs}}}
\newcommand{\ysz}{\ensuremath{y_{\rm sz}}}
\newcommand{\ywhim}{\ensuremath{y_{\rm whim}}}
\newcommand{\hatysz}{\ensuremath{\hat{y}_{\rm sz}}}
\newcommand{\chisq}{\ensuremath{\chi^2_{{\rm CO} - y_{\rm sz}}}}
\def\arcm{\ifmmode {^{\scriptstyle\prime}}
          \else $^{\scriptstyle\prime}$\fi}
\newcommand{\healpix}{{\tt HEALPix}}

\newcommand{\Tanimura}{{T19}}
\newcommand{\Graff}{{G19}}

\newcommand{\kB}{\ensuremath{k_{\rm B}}}
\newcommand{\Ne}{\ensuremath{n_{\rm e}}}
\newcommand{\Nez}{\ensuremath{n_{\rm e}^z}}
\newcommand{\barNe}{\ensuremath{\bar{n}_{\rm e}}}
\newcommand{\Te}{\ensuremath{T_{\rm e}}}
\newcommand{\me}{\ensuremath{m_{\rm e}}}
\newcommand{\sigT}{\ensuremath{\sigma_{\rm T}}}
\newcommand{\tauT}{\ensuremath{\tau_{\rm T}}}
\newcommand{\id}{\ensuremath{{\rm d}}}

\newcommand{\bs}[1]{{\color{blue} #1}}

\title{Detection of WHIM in the
  \planck\footnote{Based on observations obtained with Planck
    (http://www.esa.int/Planck), an ESA science mission with instruments
    and contributions directly funded by ESA Member States, NASA, and
    Canada.} data using {\it Stack First} approach}

\author[a]{Baibhav Singari,}
\author[a]{Tuhin Ghosh}
\author[b]{and Rishi Khatri}
\affiliation[a]{School of Physical Sciences, National Institute of Science Education and Research, HBNI, Jatni 752050, Odisha, India}
\affiliation[b]{Department of Theoretical Physics, Tata Institute of Fundamental Research, Homi Bhaba Road, Mumbai 400005, India}
\emailAdd{baibhav.singari@niser.ac.in, tghosh@niser.ac.in, khatri@theory.tifr.res.in}

\abstract{We detect the diffuse thermal Sunyaev-Zeldovich (tSZ) effect from
  the gas filaments between the Luminous Red Galaxy (LRG) pairs using a new
  approach relying on stacking the individual frequency maps. We apply and
  demonstrate our method on  $88000$ LRG pairs in the SDSS DR12 catalogue selected with an improved selection criterion that ensures minimal contamination by the Galactic CO emission as well as the tSZ signal from the clusters of galaxies. We first stack the \planck\ channel maps and then perform the Internal Linear Combination method to extract the diffuse \ysz\ signal. Our {\it Stack First} approach makes the component separation a lot easier as the stacking greatly suppresses the noise and CMB contributions while the dust foreground becomes homogeneous in spectral-domain across the stacked patch. Thus one component, the CMB, is removed while the rest of the foregrounds are made simpler even before component separation algorithm is applied. We obtain the WHIM signal of $\ywhim=(3.78\pm 0.37)\times 10^{-8}$ in the gas filaments, accounting for the electron overdensity of $\sim 13$. We estimate the detection significance to be $\gtrsim 10.2\sigma$. This excess \ysz\ signal is tracing the warm-hot intergalactic medium and it could account for most of the missing baryons of the Universe. We show that the {\it Stack First} approach is more robust to systematics and produces a cleaner signal compared to the  methods relying on stacking the $y$-maps to detect weak tSZ signal currently being used by the cosmology community.}

\date{\today}

\begin{document}
\maketitle

\section{Introduction}

According to the standard $\Lambda$CDM model of Cosmology, our Universe is
composed of approximately $5\,\%$ baryonic matter with the rest $95\,\%$ of
the total energy density in the
form of dark matter and dark energy. The current most precise  measurement of
the baryonic energy
density parameter, $\Omega_b h^2 = 0.02225 \pm 0.00016$, where
$h=H_0/(100~{\rm km/s/Mpc})$
and $H_0$ is the Hubble constant, is derived from the cosmic microwave
background (CMB) measurements \citep{Plank_Colab_cosmopara} and thus tells
us the amount of baryons present at the time of recombination at redshift
$z\approx 1100$ \citep{zks68,peebles68}. However, observations of the low redshift Universe
show that the baryon fraction today falls below the expected universal
value from CMB for
almost all regions (except for the massive haloes) \citep{McGaugh_2009}.
It has been known for sometime now that 
almost all
of the  baryons at high redshifts ($z\gtrsim2$) are accounted for in the
Lyman-$\alpha$ absorption forest \citep{Weinsberg_1997}. In contrast,  at
low redshifts  ($z\lesssim2$) we see that even after accounting for the
  baryons in stars, galaxies, Lyman-$\alpha$ forest gas along with broad
  Lyman-$\alpha$ and \OVI\ absorbers, and  hot gas in clusters of
  galaxies,  almost half of the baryons are still missing
  \citep{Shull_2012}.  This apparent
discrepancy between the direct observations spanning the electromagnetic
spectrum from radio to X-rays and the predicted baryonic mass in the
standard model of cosmology and the galaxy formation theories need to be
resolved by locating  `\textit{the missing baryons}'.

The gravitational instability of small initial Gaussian density
fluctuations results in anisotropic collapse \citep{Zeldovich_1970,shz1989} forming
sheets (Zeldovich pancakes) and filaments that make up a web like structure, the
cosmic web
\citep{ks1983,defw1985,gh1989,2df_survey,gott2005,mil2005}. Galaxies and
galaxy clusters, embedded in the knots of the
web (also known as dark matter haloes), are therefore connected by large-scale
filamentary  structures. It has been long known from simulations that
  a large fraction of the baryons are in the seemingly empty regions of the
  universe, i.e. outside the gravitationally bound haloes \citep{cen}. The
  haloes are highly overdense regions, but there are regions which are
  mildly overdense but span a much larger volume. These relatively low
  density and high volume spanning regions of sheets and filaments could be
  a rich reservoir of the missing baryons as they go undetected by the
  conventional methods.  The gas in these filaments or the intergalactic medium (IGM) have a
density of the order of ten times the mean baryon density and temperatures
between $10^5 - 10^7$\,K.  Hydrodynamical simulations suggest that this  warm hot
intergalactic medium (WHIM) could contain  $30-50\,\%$
of all baryons today \citep{Cen_2001,Dave:2001}, even 
  though the 
filaments occupy only  $6\,\%$ of the total volume \citep{Cautan_2014}. The
high ionization degree would
have prevented these baryons from being detected in absorption line surveys in
the radio and optical bands
while the low density and temperature would have prevented them from being
detected in either emission lines or in the X-ray surveys targeting thermal
X-ray emission, making them an ideal candidate for the
{\it missing baryons}.  The efforts for the detection of these missing
  baryons have been ongoing. Most of the campaigns targeting WHIM have focused on the detection of hot gas using X-rays from individual filaments \citep{xrayfail} or from the absorption spectra of quasars \citep{Shull_2012}. These methods are however able to probe only a part of
the phase space of WHIM 
  leaving about $\sim 30\%$ of the baryons still unobserved \citep{Shull_2012, Nicastro:2017}. There has been a recent work on X-ray detection of filamentary structures near cluster Abell 2744 \cite{Eckert:2015}, albeit their work probes the hotter and denser end of the WHIM, and quote a small fraction of baryons in that state.
Recent observations of OVII absorption lines in the
  X-ray spectra of a $z=0.48$ blazar provide some evidence for a significant
  fraction of baryons to be present in $\sim 10^6\,{\rm K}$ gas at $z\sim
  0.4$ \cite{nkk2018} leading to the claim by the authors that the missing baryons have been found,
  albeit in just two systems very close to a single blazar. There have
    been detections of kinetic Sunyaev Zeldovich (kSZ) effect from the
    baryons in the dark matter halos (galaxies and clusters of galaxies)  \cite{Hernandez:2015, Hill:2016, Schann:2016, Bernardis:2017}, which is
    sensitive to peculiar motion of the baryons with non zero
    velocities.  
   Recently, it was shown that the cross-correlation of angular
   fluctuations of galaxy redshifts with the kSZ effect in CMB temperature
   maps \cite{Jonas_2019} can be  sensitive to  baryons in
    regions with  overdensities consistent with those of the filaments and
    sheets in the cosmic web.

The elastic scattering of hot free electrons in the WHIM with the CMB photons
 boosts the energy of the CMB photons resulting
in a characteristic spectral distortion of the CMB, the thermal Sunyaev
Zeldovich (tSZ) effect \citep{sz1969}. The tSZ effect provides a way to 
study WHIM through multifrequency experiments, such as \planck, which can
separate the tSZ effect from the CMB and foreground emissions \citep{Planck-A22:2016,hs2014,Khatri:2016}.  The magnitude
of tSZ distortion, denoted by \ysz, is a
function of both the gas density and the temperature of the medium
\citep{Sunyaev:1972} and is given by (using \planck\ 2018 cosmological parameters
\cite{planck2018} and fully ionized primordial gas)
\begin{align}
\ysz &= \int \id s \Ne \sigT \frac{\kB
\Te}{\me c^2}\nonumber\\
 &\approx \tauT \frac{\kB
\Te}{\me c^2}\nonumber\\
&\approx 7.6\times
10^{-8}\left(\frac{\delta}{10}\right)\left(\frac{\Te}{10^7~{\rm
      K}}\right)\left(\frac{r}{10~{\rm Mpc}}\right),\label{Eq:ysz}
\end{align}
 where $\Ne$ is the free electron number density, $\Te$ is
the electron temperature, $\sigT$ is the Thomson cross section, $\kB$ is
the Boltzmann constant, $\me$ is the mass of electron, $c$ is the speed of
light, the integral is over the line of sight distance, $s$, through the
WHIM, $\delta=\rho/\rho_{\mathrm{b}}$ is the overdensity, $\rho$ is the filament
baryon density,  $\rho_{\mathrm{b}}$ is the average baryon density, $r$ is the
length of filament along the line of sight and $\tauT$ is the Thomson optical depth through the WHIM or the
filament along the line of sight. If we take the filament baryon density to be
$10\times$ average baryon density  ($\delta=10$), we get an optical depth of  $\tauT\sim
4.5\times 10^{-6}/{\rm Mpc}$ at $z=0$. For a temperature of $\Te\sim 10^{7}~{\rm K}$,
 we will get a tSZ
signal of $\sim 7.6\times 10^{-8}$ or a Rayleigh-Jeans temperature decrement
of $2\ysz\sim 0.1~{\rm \mu K}$  after integrating over $r=10$ Mpc along the line of sight. This signal is much smaller than the
noise in the current best CMB experiments, and in particular much smaller compared
to the sensitivity of \planck. Therefore, it is not possible at present to
detect the individual filaments. We can however beat down the noise by
stacking hundreds of thousands of filaments, improving the signal to noise
ratio, $S/N$, by a factor of hundreds. The \ysz\ signal from the WHIM in the
stacked objects would be detectable in the \planck\ data if we can remove the
contamination from the CMB as well as Galactic foregrounds with the same
accuracy. \\

The approach of stacking to improve the $S/N$ has been used  previously  to detect the faint tSZ signatures. The stacking of the tSZ signal in the maps released by the \planck\
	collaboration \citep{Planck-A22:2016} on the positions of known galaxy pairs of massive luminous red
	galaxy (LRGs) \citep[][hereafter T19]{Tanimura:2019} as well as constant mass (CMASS) galaxy
	samples 
	\citep[][hereafter G19]{Graaff:2019} from the Sloan Digital Sky Survey (SDSS) has been
	performed in an effort to find the missing baryons. In this
	technique, the selected close galaxy pairs within a certain radial and
	tangential distance are stacked up coherently in the \ysz\ map
        created from the Planck data by a component separation algorithm. G19
        claim a detection of around $\sim 11\%$ of the  baryons with
        $2.9\sigma$ detection of the tSZ signal, 
        leaving  $\sim 18 \%$ of the baryons still unaccounted for. A
        stacking of tSZ maps around superclusters was done in
        \cite{Tanimura_19_intercluster}, in which the authors claim a
        detection of $17\%$ of missing baryons in the intercluser gas in
        superclusters with the  tSZ effect  detected at $2.5\sigma$ significance.\\

\begin{figure}
\centering
\begin{tabular}{cc}
\includegraphics[width=0.9\linewidth]{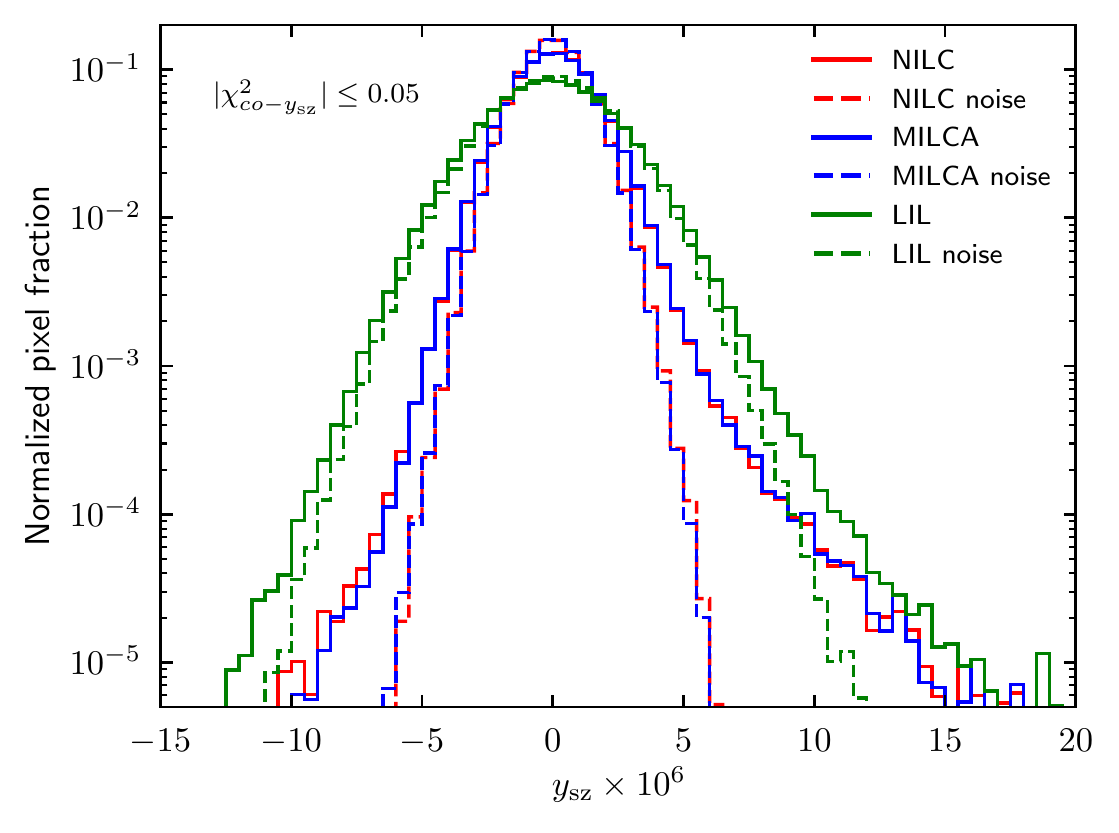} 
\end{tabular}	
\caption{The normalized distribution of the \ysz\ maps in the region
  between the LRG pairs selected as described in Sec.~\ref{sec:data} in
  MILCA, NILC and LIL \ysz\ maps. The local average background around the
  LRG pairs is subtracted to get
the correct zero level for each galaxy pair in the \ysz\ maps.}
\label{fig:Comparision_diff_maps}
\end{figure}

We present a new algorithm for detection of WHIM through tSZ effect in
 multifrequency  CMB data sets. Although the individual steps of our algorithm are similar to
  T19 and G19, the order in which the steps are performed is completely
  new. It turns out that just reordering the steps 
  makes a huge difference.  The main essence of our algorithm is to {\it first
  stack} the individual frequency maps at source locations and then
extract the \ysz\ component from the multifrequency stacked data using
standard component separation algorithms. Our new algorithm is motivated
and described in 
Sec.~\ref{sec:algo}. In Sec.~\ref{sec:data}, we introduce the \planck\ data products and the sky masks used in this paper.  Section~\ref{sec:analysis} introduces the main data analysis part of the paper to extract the \ysz\ signal at the location of LRG pairs from the stacked \planck\ maps. The modelling of \ysz\ signal expected from individual halo contribution and its subtraction from the total \ysz\ signal to see the signature of WHIM in the filament region is discussed in Sec.~\ref{sec:excess_signal}. In Sec.~\ref{sec:null_tests}, we discuss the null test and the error estimate of the excess \ysz\ signal. Finally we present our conclusions in Sec.~\ref{sec:conclusion}.

\section{A new algorithm to detect weak tSZ signals in \planck\ data}\label{sec:algo}

In the previous attempts at detecting the WHIM through tSZ effect \cite{Tanimura:2019,Graaff:2019}, there is
ambiguity as to what might be the true signal.  In the conventional method,
 stacking of sources is done  on
preprocessed publicly available \planck\ \ysz\ maps obtained from Needlet
Internal Linear Combination (NILC) \citep{Remazeilles:2011} and Modified
Internal Linear Combination Algorithm (MILCA) \citep{Hurier:2013} 
algorithms. The residual contamination by the other foreground emissions
(dust, CO, free-free, synchrotron and CMB leakage) after component separation in the
\ysz\ maps is much larger compared to the signal we are interested in
\citep{AHD2015,Khatri:2016}.  Thus, when we stack a large number of galaxy
pairs, there will be some cancellation between the positive and negative
contamination leaving a net residual systematic which can be either
positive or negative. This can be seen in the  probability distribution
functions (PDFs) of \ysz\ in NILC, MILCA and Linearized Iterative
Least-squares (LIL) maps in Fig.~\ref{fig:Comparision_diff_maps} for the
pixels which lie in-between the galaxies in a galaxy pair. The
  selection procedure is  explained in
the next section. We see that there is significant positive as well as
negative excess over the Gaussian noise. The negative excess is
contamination while the positive excess is contamination + \ysz\
signal. In particular, the contamination signal is more than a factor
  of 100 larger compared to the \ysz\ signal we are interested in. 
There is no guarantee that the positive contamination is equal to
 the negative contamination, and there would be large unknown systematic bias in
the \ysz\ signal obtained in this way.

We propose a new method of extracting the \ysz\ signal from the filaments
connecting the galaxy pairs. We
begin by first stacking the individual \planck\ frequency maps at the
positions of the LRG pairs. We  then perform the component separation using
Internal Linear Combination (ILC) \citep{tegmark1996,tegmark1998,Bennett:2003} algorithm on the stacked frequency maps to extract out the \ysz\
signal. By doing {\it stacking first} and component separation later,  we achieve
a number of  advantages over the conventional method of stacking the \ysz\
map  \citep{Tanimura:2019,Graaff:2019}:
\begin{enumerate}
\item We suppress the instrumental noise before doing ILC. Thus even the noisy 70 GHz and
  100 GHz channel are utilized efficiently. In conventional method, these
  channels are down-weighted as the ILC tries to strike a compromise
  between reducing noise and reducing foregrounds in the final map.  
\item The Gaussian random CMB fluctuations, uncorrelated with the positions
  of the galaxies, are suppressed. Thus we have one less component even
  before we begin the ILC. 
\item Since Galactic foregrounds are also uncorrelated with
  the galaxy positions, by stacking the frequency maps, we are
  homogenizing the foregrounds in amplitude as well as in spectral shape
  across our stacked  patch. That is, in every pixel in the final stacked map
  after stacking  patches of interest from different parts of the sky,
 we should expect a sum of foreground contamination from a large number of
  sources, essentially sampling the whole foreground parameter space. Every
  pixel should end up with a very similar foreground contribution,
  effectively summing the complicated foregrounds comprising of many
  different components varying across the sky to a single foreground
  component across our patch.
\end{enumerate} 
The full power of ILC is therefore concentrated in eliminating the
foregrounds, as noise and CMB is eliminated in the pre-ILC stage. The
foreground spatial structure is also simplified by stacking the frequency
maps. In particular, the dust emission has spectrally smooth behaviour in
the stacked frequency images, with negligible variation across the image,
 as a result of averaging.  In our approach, since
we are suppressing noise before doing ILC, we can work at higher resolution
compared to the conventional method. We will demonstrate this by doing
analysis at $8'$ resolution although most of our results would be derived
at $10'$ resolution.

\section{Data and masks}\label{sec:data}

\subsection{SDSS and \planck\ data}
\begin{figure}
\centering
\includegraphics[width=0.5\linewidth,angle=90]{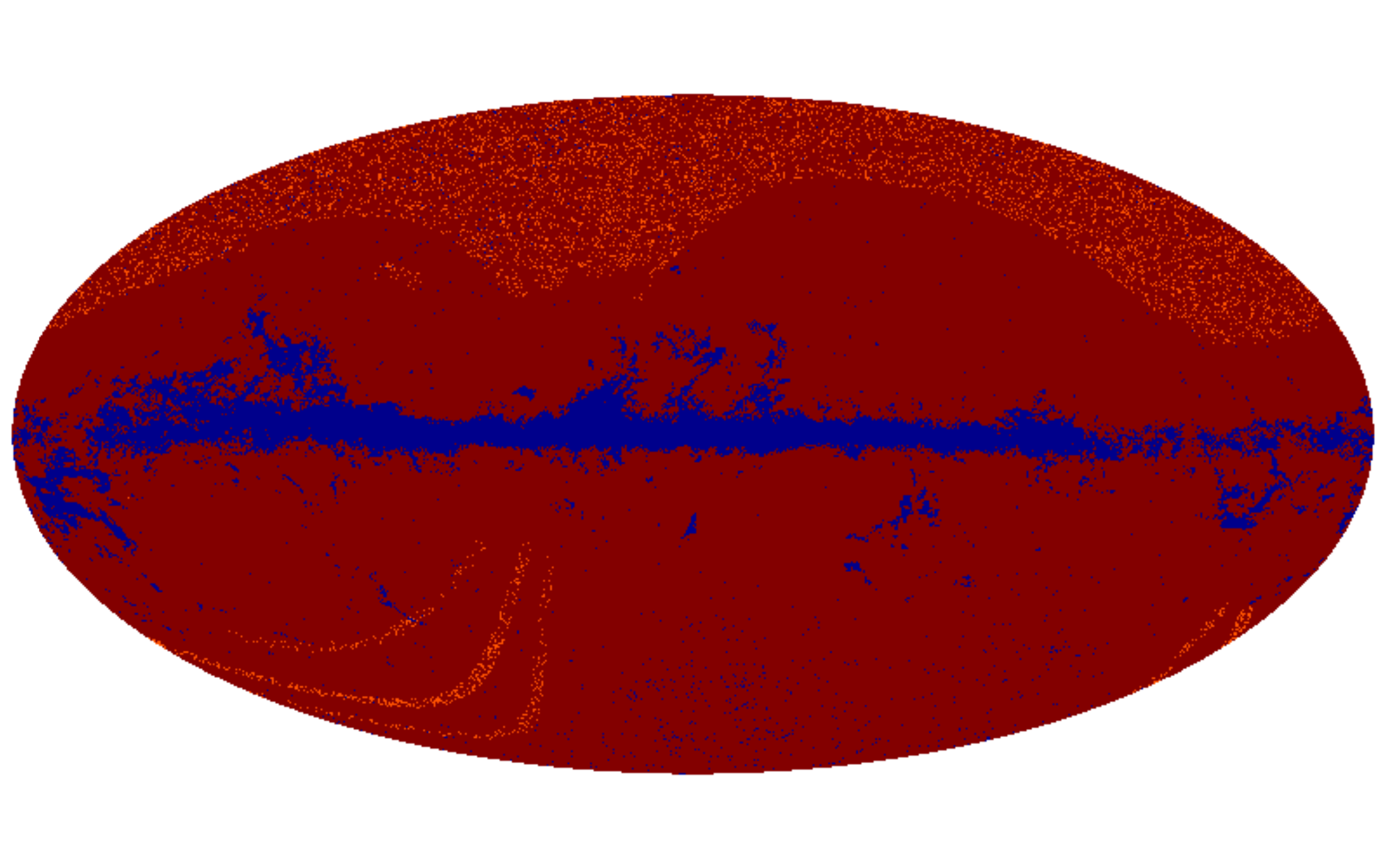} 
\caption{The eK86 mask used in our analysis.  The orange points over
  the eK86 mask represents the LRG locations in the SDSS12
  survey.}
\label{fig:mask86}
\end{figure}
We use SDSS data release 12 (DR12) with the same criteria as T19 to make a
catalog of LRGs with stellar mass $\Mstar > 10^{11.3} \Msolar$
\citep{Prakash:2016}. We take the stellar mass estimate based on a
principal component analysis method from \citep{Chen:2012}. Next, we
construct a sample of LRG pairs with the  radial distance between the
galaxies of a pair  $\le\,6h^{-1}$\,Mpc and the tangential distance in the
range $6 - 10h^{-1}$\,Mpc. We adopt a $\Lambda$CDM cosmology with
$\Omega_{\rm m}=0.3$, $\Omega_{\Lambda}=0.7$, and $H_0=70\,\text{km
  s}^{-1}\Mpc$ for calculation of the comoving distances from the redshift
($z$) information. If two or more LRG pairs fall within 30\arcm\ in the projected
sky coordinates (Galactic latitude and longitude), then we only keep the higher average mass LRG pairs in the sample. We find roughly $161000$ LRG pairs satisfying both of
the distance  criteria in the SDSS DR12 sample. The angular separation
between the selected LRG pairs lies between 19\arcm\ and 203\arcm\ .

We will use the \planck\ 2015 intensity maps from 70 to 857 GHz and IRIS 100 $\mu$m (or 3000 GHz) map \citep{MAMD-IRIS} for our
analysis. The temperature data has not changed significantly between 2015
and 2018 releases. We rebeam the \planck\ HFI maps to a common beam
resolution of 10\arcm\ full
width half maximum (FWHM), taking into account the effective beam
function of each map and reduce to a \healpix\ resolution of $\Nside=1024$
from the original $\Nside=2048$ to make computations faster. While smoothing to 10\arcm\ beam resolution, we only retain the scales up to $\ell_{\rm max}=4000$ for \planck\ HFI channels and $\ell_{\rm max}=2048$ for \planck\ 70 GHz LFI channel. As a validation step, we also produce \planck\ maps at a common beam resolution of 8\arcm\ FWHM with $\ell_{\rm max}=3000$ for \planck\ HFI channels and keeping 70 GHz LFI channel $\ell_{\rm max}=2048$.

We will use the MILCA and NILC \ysz\ maps  from the \planck\ Legacy
Archive\footnote[1]{\url{https://www.cosmos.esa.int/web/planck/pla}} for
comparison. The MILCA \ysz\ maps were produced using  all of the
\planck\ High Frequency Instrument (HFI) intensity maps
($100-857$\,GHz). The NILC method uses in addition the Low Frequency
Instrument (LFI) data ($30-70$)\,GHz) at large angular scales ($\ell <
300$). The angular resolution of both of these \ysz\ maps is 10\arcm\  FWHM. We downgrade the original  \ysz\ maps from
$\Nside=2048$ (pixel size=1.7\arcm) to $\Nside=1024$ (pixel size=3.4\arcm)
for computational efficiency.

\subsection{Sky masks and sample selection}
\begin{figure}
\resizebox{3.0in}{2.2in}{\includegraphics{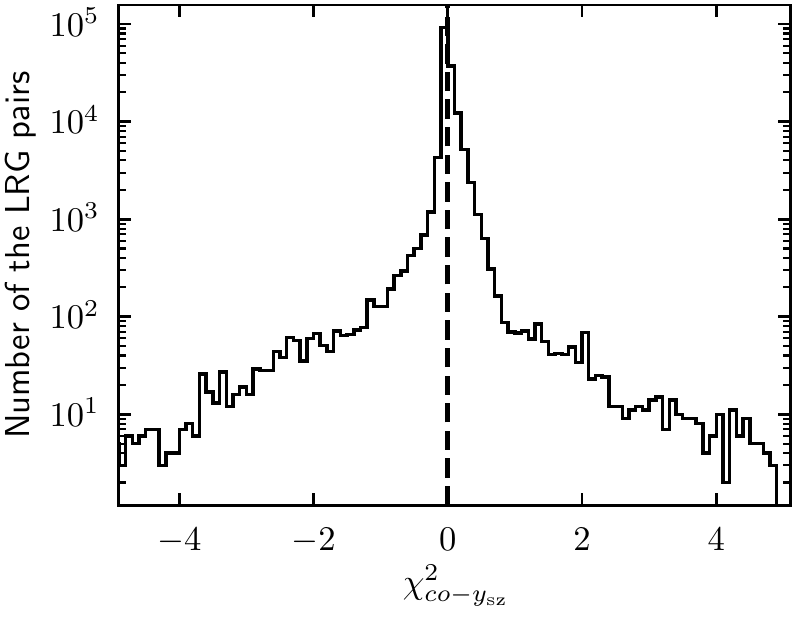}}
\hskip 0.0cm \resizebox{3.0in}{2.2in}{\includegraphics{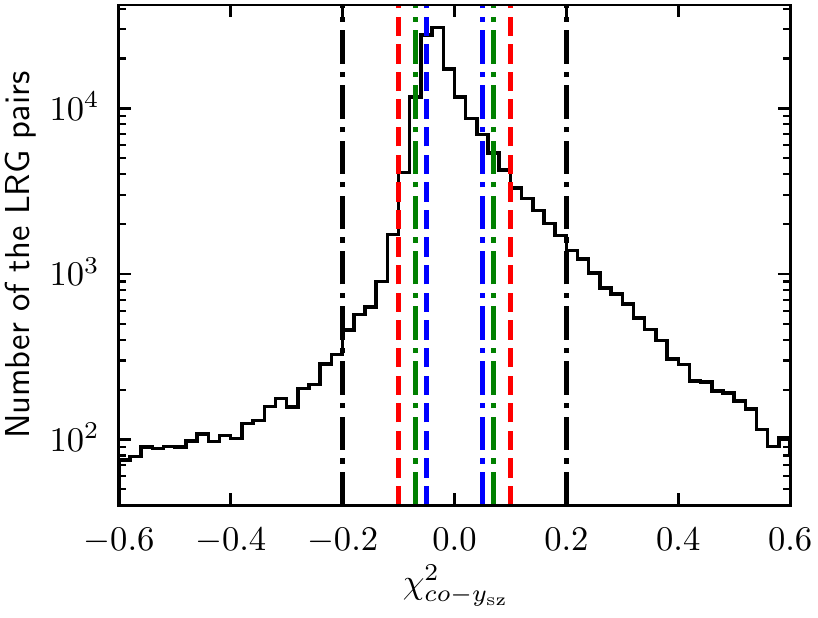}}
\caption{The $\chisq$ distribution at the location of LRG pairs. Right
  panel shows a zoomed-in version. The vertical lines represent different $\chisq$ 
  thresholds used in our stacking analysis.} 
\label{fig:chisq}
\end{figure}
We will use the sky mask obtained in \citep[][hereafter K86 mask]{Khatri:2016} specifically for
the tSZ studies with an unmasked sky fraction $\fsky=86\%$  as the
baseline. This mask specifically tries to minimize the CO line emission
contamination and also covers strong point sources. We will also use the
masks provided by the \planck\ collaboration, in particular the Galactic + point
source mask with $\fsky=48\%$ (henceforth PL48 mask) to compare with the
\Tanimura\ results. 

\begin{figure}
\resizebox{3.0in}{2.2in}{\includegraphics{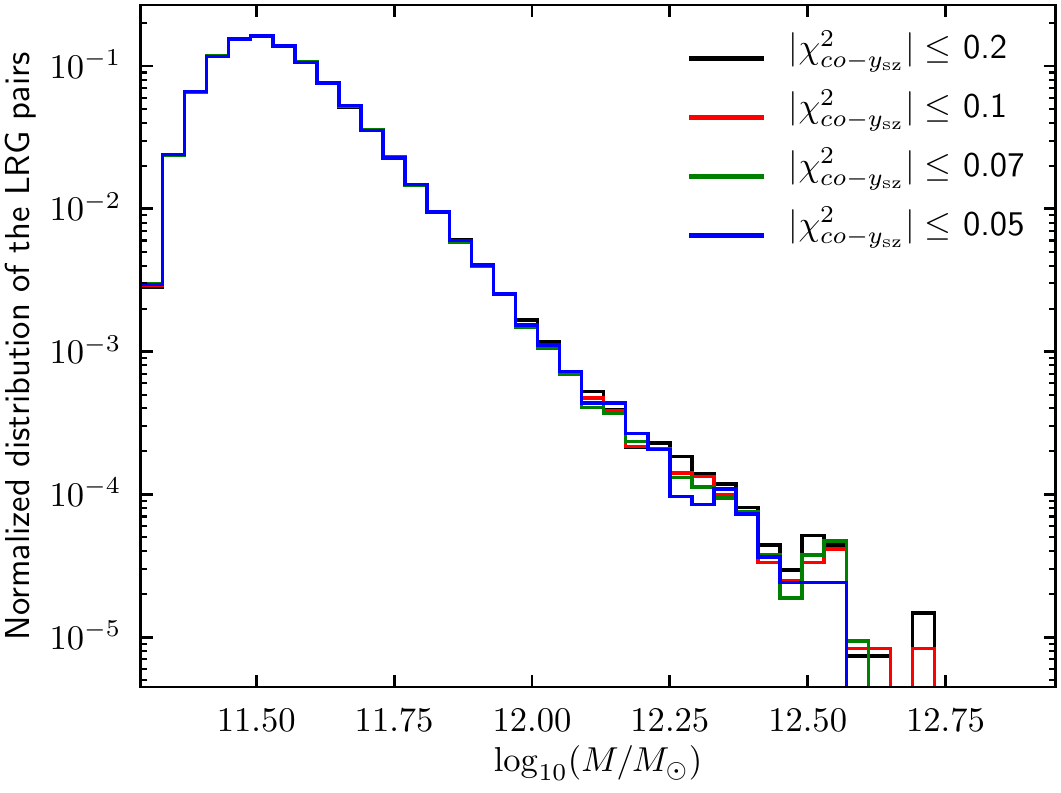}}
\hskip 0.0cm \resizebox{3.0in}{2.2in}{\includegraphics{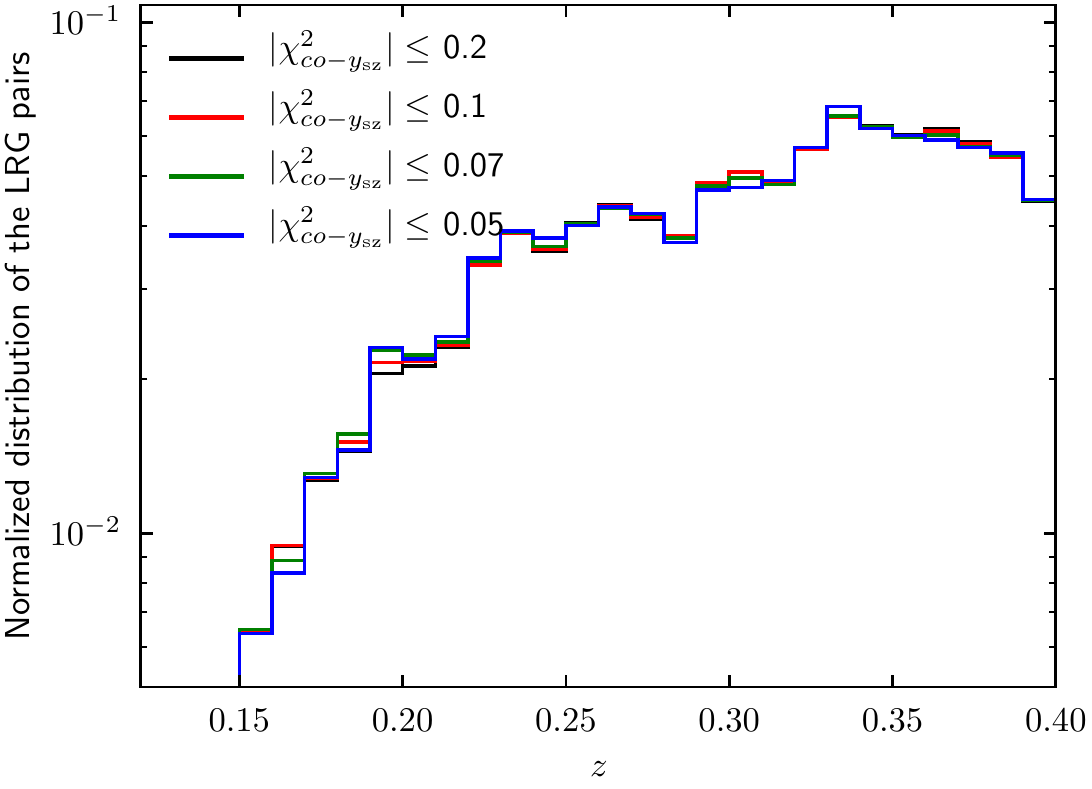}}
\caption{ \textit{Left panel}: the normalized distribution of the mean mass
  of the LRG pairs for four different \chisq\ thresholds. \textit{Right
    panel}: the normalized distribution of mean redshift distribution of
  the LRG pairs as a function of \chisq\ thresholds. }
\label{fig:redshift_mass}
\end{figure}
As we estimated in the last section, we expect the IGM in the filaments between the galaxies
to give a very weak tSZ signal, much below the noise level of the \planck\
for individual objects. Thus, in addition to the regions of strong
CO line contamination \citep{Khatri:2016}, we also want to avoid the strong tSZ
signal coming from much hotter and denser gas in the clusters of galaxies
in the foreground or background, i.e. we want to select only those  pairs of
galaxies for which, in the individual objects, the \ysz\ signal is undetectable
and we are dominated by the instrumental noise. To accomplish this, we
  use the fact  that CO emission is also a weak signal in
  the \planck\ data, of similar strength to the \ysz\ signal but with a different
  spectrum.  We can fit a model consisting of CMB + Dust + tSZ signal to the
\planck\ HFI data as well as a model consisting of CMB + Dust + CO emission,
and compare the $\chi^2$ of the two models (which have the same number of
parameters). In the regions where CO emission is stronger than the
tSZ, we will have smaller $\chi^2$ for the CO model and the difference
between the $\chi^2$ for the two models \chisq will be negative. In the
opposite case, when we have stronger tSZ signal, \chisq will be positive. We
want to avoid both these cases. We want to select galaxies such that we are
noise dominated and unable to distinguish between the two models,
i.e. \chisq$\sim 0$. These model fits were performed in \cite{Khatri:2016} and
we will use the \chisq map obtained in \cite{Khatri:2016} to further prune our galaxy
sample. For each LRG pair, we attribute a \chisq\ value by computing
  the average \chisq\ of the sky pixels that lies within  20\arcm\
  radius from the centre of LRG location. We have $\sim 99.6\%$ of our
  sample with a $|\chisq| < 5$ and $\sim 96.6\%$ of our sample with a
  $|\chisq| < 0.5$. We can thus use aggressive thresholds in $\chisq$
    removing the most contaminated galaxy pairs but still loose only a
    small fraction of the sample.

We first extend the K86 mask by masking the sky pixels where the \chisq\ values are either
highly negative or highly positive, i.e. $|\chisq|>5$. This extended
K86 mask, hereafter eK86, is shown in Fig.~\ref{fig:mask86}. This extension  masks the \planck\ detected 1653 clusters \citep{Planck-A27:2016}, SZ dominated
regions and molecular clouds from our concerned sample. If either of the
two LRGs of the pair falls in the masked region, then we exclude that LRG pair from our stacking analysis. The PDF of average \chisq for all galaxy pairs in our
sample, after applying the eK86 mask, is plotted in
Fig.~\ref{fig:chisq}. The skewness towards positive values in the
distribution of \chisq\ with eK86 mask
towards the positive side shows that in the concerned sky regions  \ysz\
signal dominates over the foreground CO emission. To get an even cleaner
sample, we further eliminate the galaxies in the tails of the PDF and
consider only those LRG pairs for stacking for which  $|$\chisq$|< 0.2$,
marked by vertical lines in Fig.~\ref{fig:chisq}. We can get cleaner
samples by further reducing the \chisq threshold. Also in order to test
that the \chisq threshold does not affect our results,  we  will consider
samples with thresholds of $|\chisq| \leq 0.2, 0.1, 0.07, \text{and} \,
0.05$ containing  144930, 128528,
113374, and 88001 LRG pairs respectively.  We show in Fig.~\ref{fig:redshift_mass}
the distributions of mean mass ($\log (M/M_{\odot})$) and mean redshift of the LRG
pairs for the four   \chisq\ thresholds. We see that the mass and redshift
distributions are insensitive to the \chisq thresholds. The \chisq
  thresholds, therefore, do not introduce any bias in our sample.

\section{Stacking analysis}\label{sec:analysis}
We use the following procedure to stack the \planck\ sky maps at each
frequency. The angular separation between the LRG pairs in our sample spans from 19\arcm\ to 203\arcm.
We first project a given LRG pair from the spherical coordinates onto a
normalized tangent plane centered at the midpoint of the line joining the
pair such that  one LRG is placed at $(-1, 0)$ and the other LRG at $(1, 0)$
\citep{Clampitt_2016}. We interpolate the tangent plane
projections of all the LRG pairs to an equal size grid. For our analysis we project the tangent plane to a grid of $301 \times 301$ pixels. The angular resolution of each pixel in the grid concerned varies from pair to pair. The two LRGs are always placed 50 pixels apart in the grid, the pixel resolution varies from $\sim 0.4\arcm$ for the pair with least angular separation to $\sim 4.3\arcm$ for the largest. We then stack on the equal sized grids. We interchange the one LRG location from $(-1, 0)$  to $(1,0)$ and other vice versa to produce the symmetric stacked signal on both LRG positions. 
We perform the stacking  for the four  \chisq\ selected
samples as discussed in the previous section  for all of the \planck\ HFI
maps and the 70 GHz LFI map at  8\arcm\ and 10\arcm\ FWHM resolutions. We
note that these are slightly finer compared to the native 70 GHz channel
resolution and therefore the noise in 70 GHz maps gets boosted. However,
the noise is suppressed again when we stack and we can thus hope that the
70 GHz channel (as well as 100 GHz channel at 8\arcm) will contribute to the
\ysz\ signal.  

\begin{figure}[!ptbh]
\centering
\includegraphics[width=0.8\linewidth]{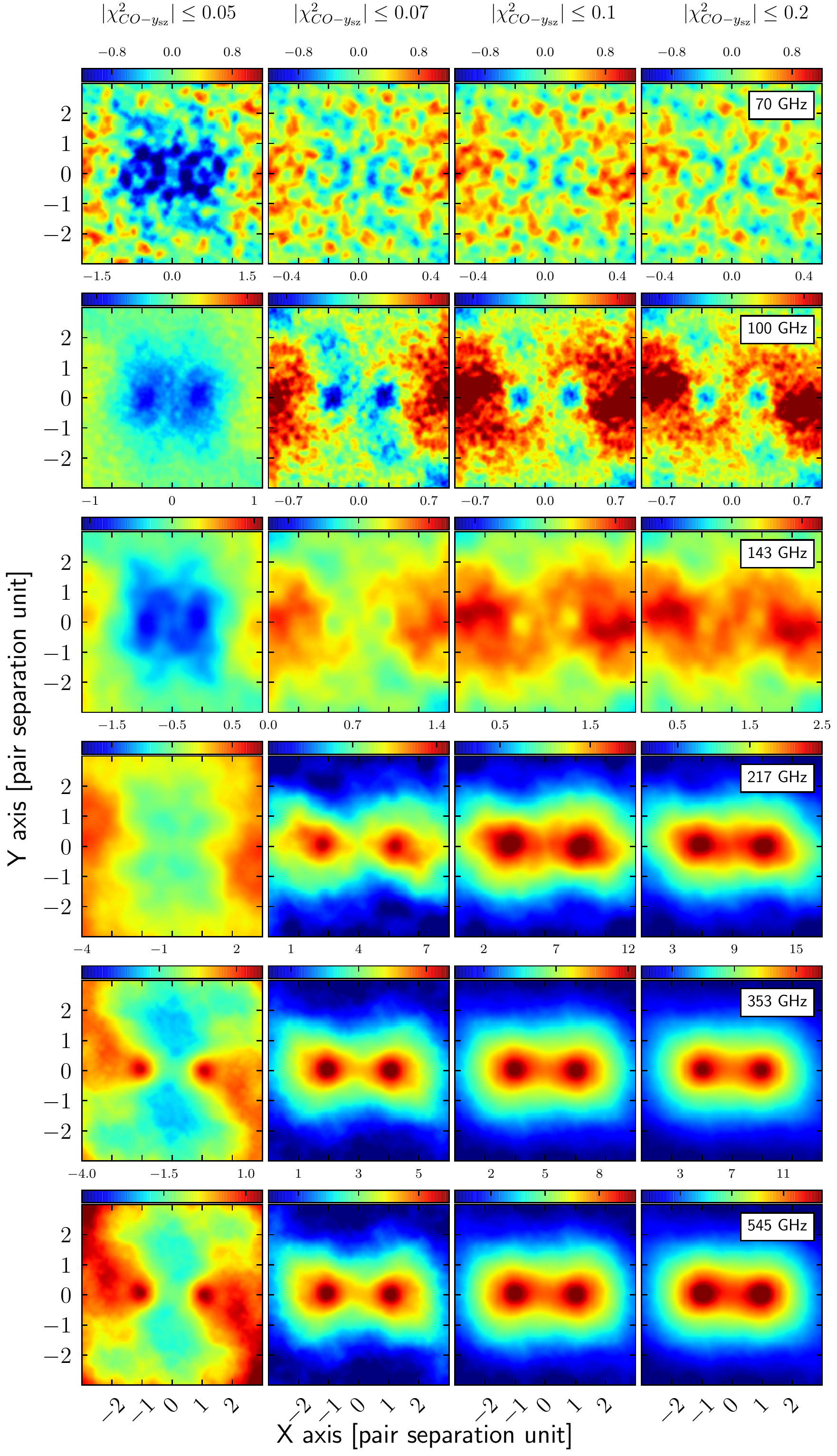}
\caption{The stacked \planck\ maps smoothed at common 10\arcm\ FWHM beam resolution for different \chisq\ thresholds:  $|\chisq| \leq 0.05$ (\textit{column 1}),  $\leq 0.07$ (\textit{column 2}), $ \leq 0.1$ (\textit{column 3}),  and $ \leq 0.2$ (\textit{column 4}). The top to bottom rows represent different \planck\ channels starting from 70 GHz  (\textit{top row}) to 545 GHz  (\textit{bottom row}). All the \planck\ maps from 70 to 353 GHz are expressed in $\mu \text{K}_{\rm cmb}$ units and 545 GHz map is in kJy/sr. The zero level of the stacked \planck\ channels is adjusted such a way that the stacked signal in the range $-3 < \text{X} <3$ and $ 0.05 < \text{Y} < 0.05$ is set to 0.}
\label{fig:stacked_point07_10}
\end{figure}

We show the stacked \planck\ frequency maps in
Fig.~\ref{fig:stacked_point07_10}. The stacked maps show an increase in foreground
contamination as we increase the $|\chisq|$ thresholds from
0.05 to  0.2.  We see the unmistakable signatures of the hot gas in the
filament region due to the tSZ effect in Fig.~\ref{fig:stacked_point07_10},
i.e. a negative signal at 143 GHz and lower frequencies with respect to the
average  ambient background around the galaxies and a positive signal for
$\nu > 217$\,GHz. This signature becomes slightly  less prominent for
higher \chisq thresholds. We will be using our cleanest sample with
  $|\chisq|<0.05$
 for our baseline analysis. The signal at the position of galaxies
  is dominated by the radio and infrared emission from the galaxies
  themselves. We expect this galactic contamination to become subdominant
  as we go away from the centers of the galaxies to the intergalactic
  medium. We fit the modified blackbody spectrum to every pixel in the 
  stacked image from 217 to 3000 GHz. The dust temperature in the fit is fixed
  to 18 K. The fitted dust amplitude normalised at 353 GHz has the same morphology as the 
  353 GHz stacked \planck\ map. After taking into account the color correction factors due
  to the \planck\ bandpasses, the fitted dust spectral indices are close to 1.4 over the entire stacked patch. 
  We can conclude that the dust emission has spectrally smooth behaviour as a result of averaging
  of dust spectral energy distribution over the different line of sights. 
  We will   need to remove the galactic contamination from the LRGs
  themselves to get an
  unbiased estimate of the \ysz\ signal from the intergalactic medium.

\afterpage{\clearpage}

\subsection{Blind component separation}\label{sec:ilc}
 We will use the ILC method \citep{tegmark1996,tegmark1998,Bennett:2003}  on the stacked \planck\ maps to extract the \ysz\ signal.
The ILC is a blind component separation method used to extract the signal
of interest, whose spectrum is known, from multifrequency observations
without assuming anything about the frequency dependence of unwanted
foreground contamination. It has been used extensively in the CMB data
analysis in the past to extract the CMB signal from multifrequency
\textit{Wilkinson Microwave Anisotropy Probe (WMAP)} sky observations. The
ILC method can be  applied over distinct regions of the sky in pixel space 
\citep{Bennett:2003, Eriksen:2004}, domains in harmonic space
\citep{Tegmark:2003}, or domains in needlet space
\citep{Basak_NILC_2012}. The NILC and MILCA component separation methods
used by the \planck\ collaboration to extract the \ysz\ signal from the
multifrequency \planck\ maps are also based on the  ILC method with
additional constraints \citep{Remazeilles:2011, Hurier:2013}. 

The ILC is a multifrequency linear filter that minimizes the variance of
the reconstructed \ysz\ map. We assume the stacked maps ($x$) in each
\planck\ channels as a superposition of \ysz\ signal, foreground ($f$) and
 noise ($n$), written as $x_i(p)= a_i \ysz(p)+ f_i(p)+n_i(p)$, where the
index $p$ labels the pixels in the stacked  map. The coefficients $a_i$ contains the
relative strength of the \ysz\ signal in the different \planck\
channels. The ILC solution for the tSZ signal, $\hatysz (p)$  is given by
$\hatysz(p)=\sum_i w_i x_i(p)$.  The weights, $w_i$, are found by
minimizing the variance of $\hatysz(p)$ subjected to the constraint that
the \ysz\ signal is preserved, i.e. $\sum_i a_i w_i=1$ \citep{Remazeilles:2011, Hurier:2013}.
\begin{figure}
\includegraphics[width=1.0\linewidth]{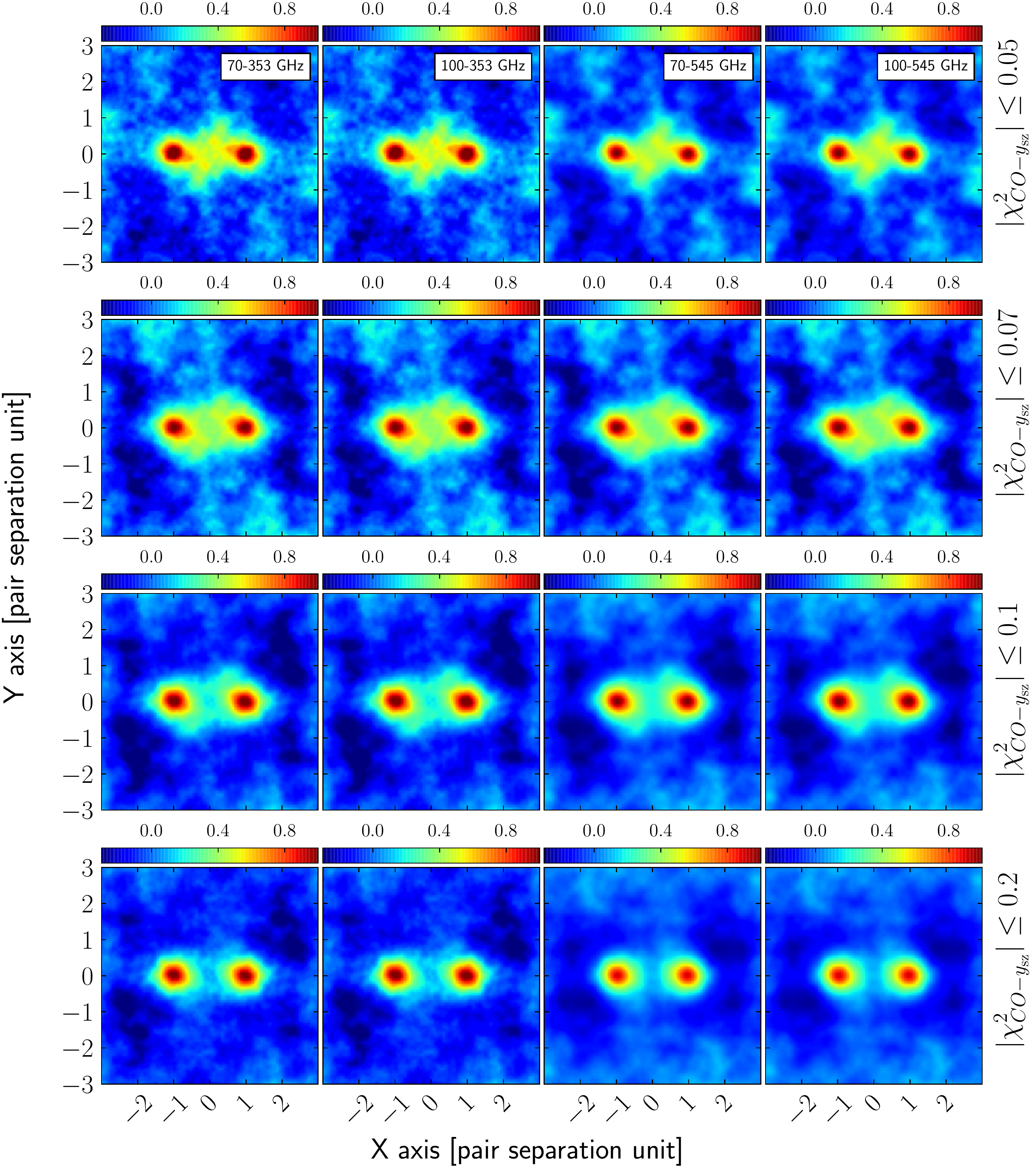}
\caption{The stacked ILC \hatysz\ signal expressed in units of $10^{-7}$  extracted from the stacked common 10\arcm\ FWHM beam resolution \planck\ maps at $\Nside=1024$ for the four different thresholds:  $|\chisq| \leq 0.05$ (\textit{column 1}),  $\leq 0.07$ (\textit{column 2}), $ \leq 0.1$ (\textit{column 3}),  and $ \leq 0.2$ (\textit{column 4}). The top to bottom rows represent combination of different \planck\ channels used to extract the stacked ILC \ysz\ signal: $70-353$\,GHz (\textit{row 1}), $100-353$\,GHz (\textit{row 2}), $70-545$\,GHz (\textit{row 3}) and $100-545$\,GHz (\textit{row 4}). }
\label{fig:ILC_10}
\end{figure}
We use the ILC method in pixel space to reconstruct the stacked \ysz\
signal from the stacked \planck\ maps. We show the \ysz\ map for our
four LRG pair samples with different \chisq thresholds and using
different frequency channel combinations in
Fig.~\ref{fig:ILC_10}.  
 
The ILC \hatysz\ maps are quite robust w.r.t to the
  changing \chisq thresholds as well as the number of frequency channels. We will use the
reconstructed ILC map derived from the frequency range $70-545$\,GHz for our fiducial analysis. 
The ILC weights obtained from our analysis with K86 mask and
    selection criterion $\chisq \le 0.05$ are quoted in
  Table~\ref{tab:ilc_weights} and compared with the ILC weights
    obtained for unstacked maps. We see from
Fig.~\ref{fig:ILC_10} that  removing 70 GHz and/or 545 GHz channels does not
make a significant difference, implying that the ILC has converged as far
as the number of frequency channels is concerned. The strong signal at the
galaxy positions is indicative of residual contamination
  from the emissions of the stacked galaxies themselves which needs to be
  estimated and removed.

\begin{figure}
\centering
\includegraphics[width=0.7\linewidth]{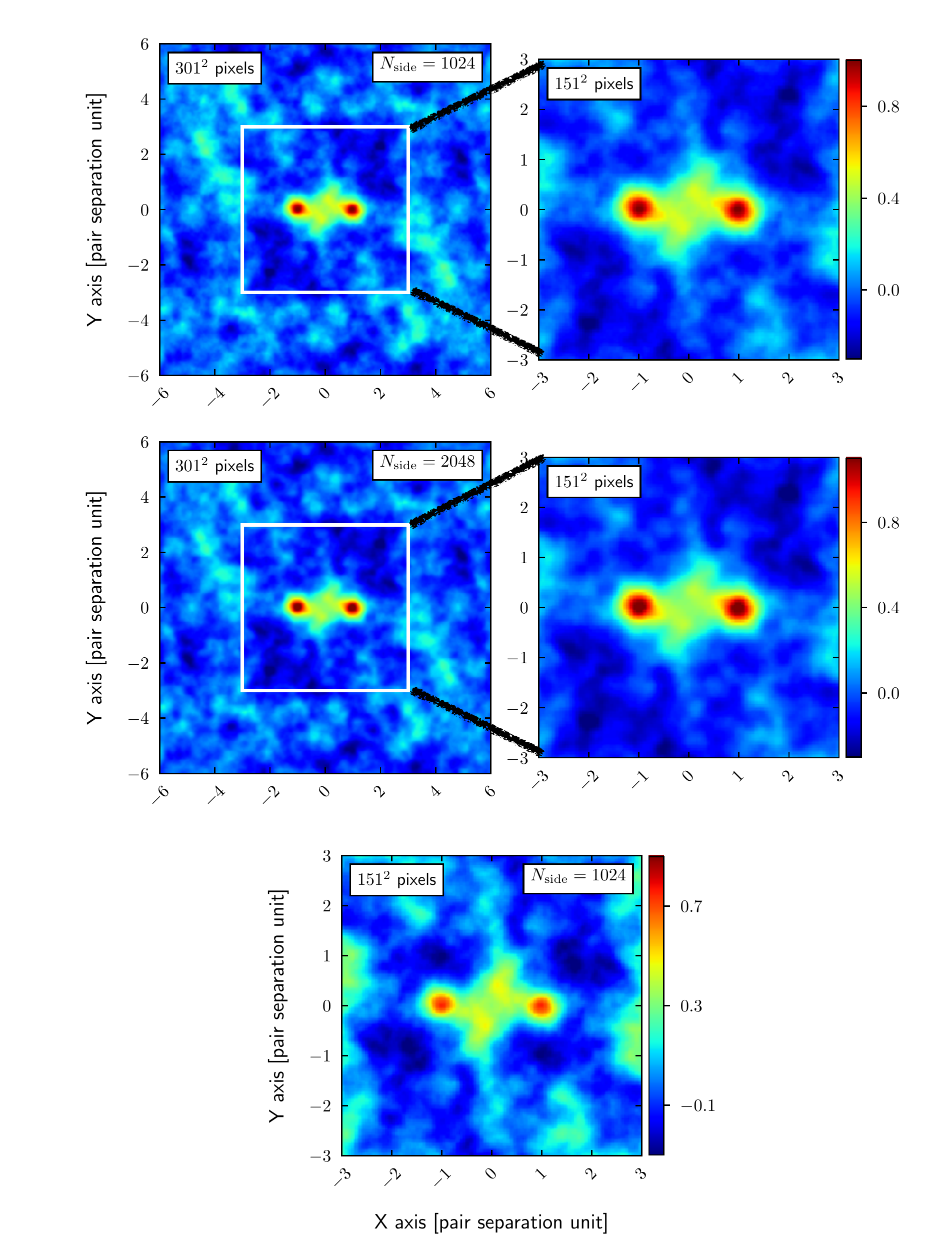}
\caption{The ILC \hatysz\ signal expressed in $10^{-7}$ units extracted from \planck\ $70-545$\,GHz channels at \Nside=1024 and a patch size of $301\times 301$ pixels (\textit{top panel}), \Nside=2048 and a patch size of $301\times 301$ pixels (\textit{middle panel}), and \Nside=1024 and a patch size of $151\times 151$ pixels (\textit{bottom panel}) for the threshold $\chisq \le 0.05$ and K86 mask. }
\label{fig:ILC_tests}
\end{figure}

Since we are doing ILC on a small number of pixels, there maybe ILC bias
coming from chance correlations between the noise, \ysz\ signal, and
foregrounds \cite{tegmark1996,dc2007,dc2009}. To check for the ILC bias we
half the size of the patch around the LRG pairs from $N\times N$
  arcmin to $N/2\times N/2$ arcmin and the number of interpolated pixels from  $301
  \times 301$ to $151 \times 151$. The results of comparison are shown in
  Fig.~\ref{fig:ILC_tests}. We also do the analysis at original \healpix\
  resolution of $\Nside=2048$. We see no significant evidence for ILC
  bias. At higher resolution of $\Nside=2048$, the signal is a little
  smoother, but otherwise consistent with $\Nside=1024$ results.

We also validate our stack first approach on simulations of realistic
  sky simulations in Appendix \ref{sec:appa} and show that our stack first
 + ILC approach recovers the WHIM tSZ signal without bias or significant
 foreground contamination. We note that since we are not interested in the
 mean background signal, we implement the standard ILC in which the cost
function that is minimized is the variance of the reconstructed $y$ map
\cite{Eriksen:2004,dc2007} i.e. the fluctuations about the mean are minimized. There are other
choices for the cost function possible. An alternative
cost function proposed in \cite{Khatri_FCILC_2019} minimizes L2 norm of the
total signal, including the average signal, especially useful when we are
interested in the large scale modes. We tried both formulations
and found that the two choices  of the cost function give consistent results,
 with the standard ILC giving a  slightly  stronger detection significance. The fact
that the two formulations give results which are similar can be taken as
additional evidence that the foregrounds are getting homogenized by
stacking as expected.

\begin{table}[h!]
\centering
\caption{The ILC weights applied to individual \planck\ stacked maps to
  reconstruct stacked ILC \hatysz\ signal over different Galactic sky masks
  used in our analysis. For comparison we also show the weights with
    the LRGs selected using PL48 mask and using 6 \planck\ channels. The last
    column shows the weights when    ILC is applied to pixels at the LRG positions in unstacked Planck
    maps. It is clear that the solution we get for ILC on stacked maps is very
    different from the one we get for the ILC on unstacked maps.}
\label{tab:ilc_weights}
\resizebox{0.8\textwidth}{!}{
\begin{tabular}{ |c|c|c|c|}
\hline
Frequency &  \multicolumn{3}{c|} {ILC weights}  \\
\cline{2-4}
in GHz & $|\chisq| < 0.05$ with K86 & PL48 mask & Unstacked pixels\\
\hline
70    & $-0.0024$ & $-0.0153$ &  $-0.0008$ \\
100  &  {    }\, 0.0182 & $-0.0857$ & $-0.0057$ \\
143  & $-0.5160$ & $-0.1158$  & $-0.2710$ \\
217  & {    }\, 0.5861& {    }\, 0.1518 & {    }\, 0.2444\\
353 & $-0.0808$ & {    }\, 0.0465  & {    }\, 0.0352\\
545 & {    }\, 0.0012& $-0.0040$ & $-0.0031$ \\
\hline
\end{tabular}
}
\end{table}

 \subsection{Component separation by parameter fitting}\label{sec:lil}
In order to use  parameter fitting to do component separation, we need an
accurate model. However stacking on frequency maps mixes different
foreground spectra, in particular mixes dust, CO and low frequency
emission. It is therefore difficult to come-up with a foreground that will
be accurate enough for our purpose. However, we can still use parameter
fitting to learn about the foregrounds and check our assumptions. In
particular, we want to check whether the foreground shape is really
homogenized over our patch by stacking and that any residual foreground
contamination is morphologically different from the SZ signal we are
interested in.
We employ  LIL parametric fitting algorithm developed in
\citep{k2015,Khatri:2016}. We fit a simple parametric model consisting of
either 
CMB + dust + tSZ or CMB + dust + CO components  to 4 HFI channels from 100~GHz to 353~GHz. We model dust by a modified blackbody
spectrum with fixed temperature $T_d=18~{\rm K}$ and fixed line ratios
for CO line contribution in different channels, following
\citep{Khatri:2016}. The
parameters to fit are CMB temperature, tSZ or CO amplitude, dust amplitude
and the dust spectral index. 

\begin{figure}
\centering
\begin{tabular}{ccc}
\includegraphics[width=1.05\linewidth]{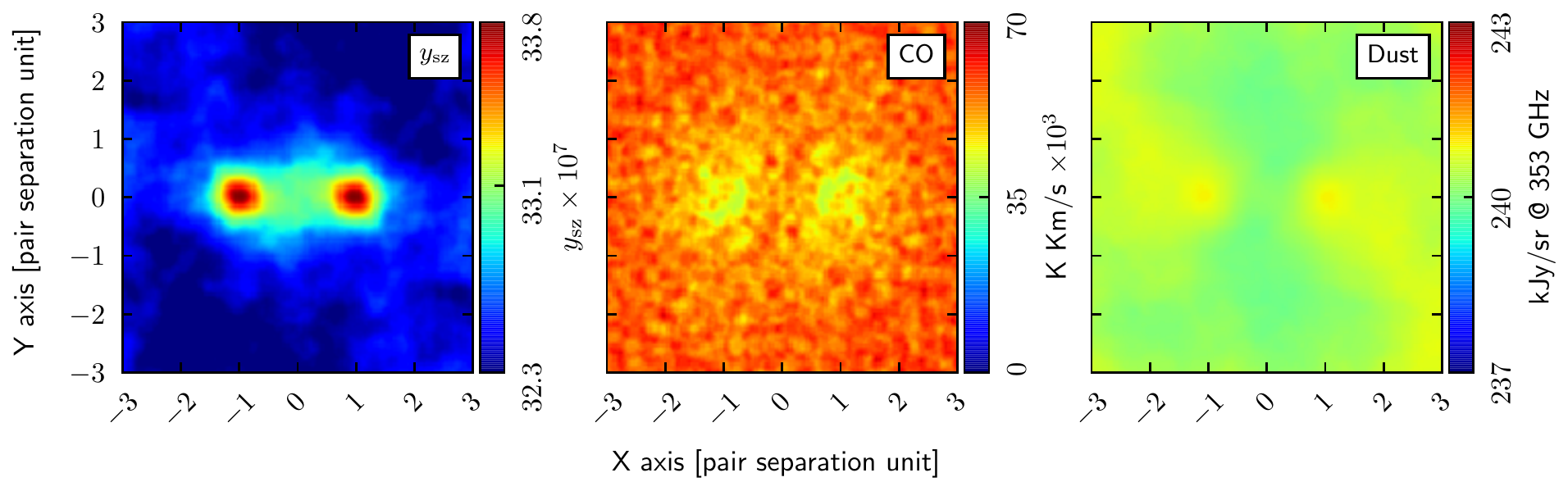} \\
\includegraphics[width=1.05\linewidth]{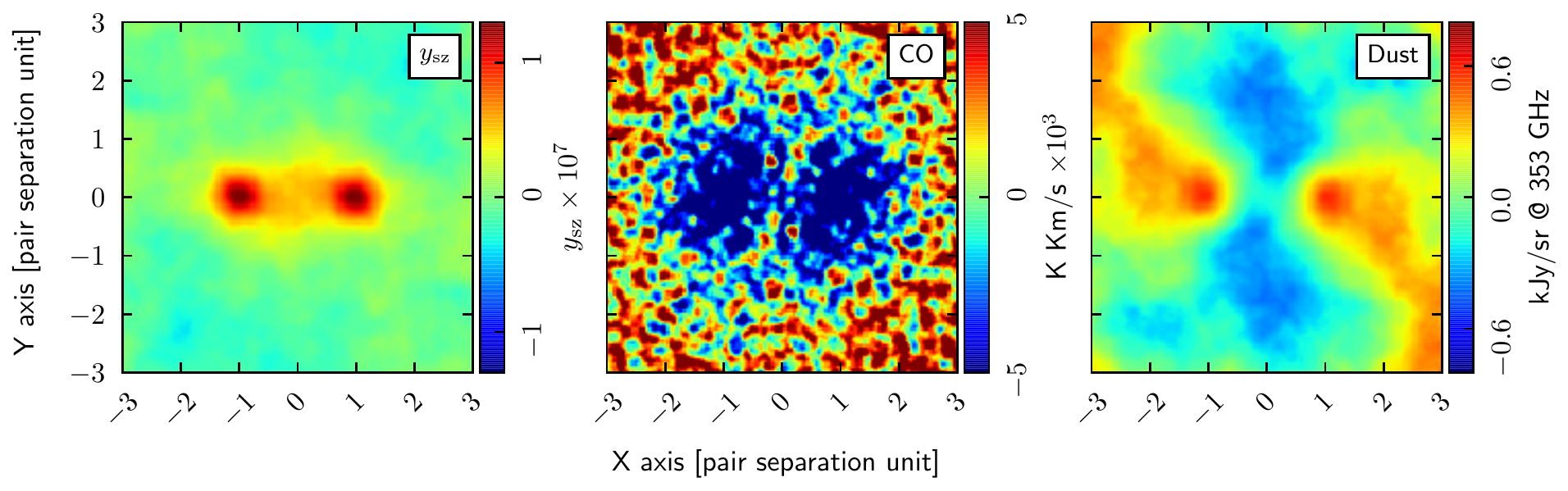}
\end{tabular}
\caption{The stacked \ysz\ signal (\textit{left panel}), CO emission
  (\textit{middle  panel}) and dust amplitude ({\it right panel}) obtained from 
three-component LIL parameter fitting. \textit{Top panel} is the fitted LIL component maps. In the \textit{bottom panel}, the average level of the \ysz\ map, CO map and the dust amplitude map is subtracted off to accentuate the visibility of the fluctuations.}
\label{fig:LIL_results}
\end{figure}

The results of the parameter fitting exercise are  shown in
Fig.~\ref{fig:LIL_results}. The top panel shows the fitted SZ signal,
  the CO signal and the dust amplitude (for SZ+dust+CMB model fit)
  signal. In the bottom panel, we have removed the average background to
  show the fluctuations. We recover the \ysz\ signal with morphology
remarkably similar to the ILC method. We also see that both the dust
  and the CO signals are quite homogeneous.  Comparing
  the top and bottom panels, we  see that the dust amplitude is almost
  homogeneous over the entire patch with 
  fluctuations of  $\lesssim 0.1\%$. For the CO signal also, the fluctuations are smaller by a
  factor of $\sim 4-5$ compared to the average signal. In particular the CO
  fluctuations are  dominated by noise and are also statistically
  homogeneous and random, apart from the small leakage from other
    components at the locations of the galaxies. In particular the
   CO signal  fluctuations are of order few$~\times 10^{-3} ~{\rm K km /s}$ which
   corresponds to few $\times 10^{-8} {\rm K_{CMB}}$ in CMB temperature units at 100
   GHz \cite[see][for conversion factors]{planckhfi,k2015}. This is approximately the amplitude of the noise
    after stacking $\sim 10^5$ galaxies in the most sensitive Planck maps
    with full sky map level sensitivity of $\sim 10^{-5}~{\rm K}$
    \cite{hfi2016,hfi2018}.

We note that CO is already a very weak contamination in
  Planck maps (see also Appendix \ref{sec:appa}). 
  We also note that while doing standard ILC \cite{Eriksen:2004,dc2007}, we subtract out the average or the monopole part of the signal. Thus $99.9\%$ of the dust and most of the CO contamination would be subtracted out even before we do the ILC. Most of the background SZ signal, seen in the top panel, will  also be suppressed.
  The small correlation in the CO map with the galaxy positions is because
  when we fit for the CO, our model does not include the SZ component. The
  SZ component will therefore contaminate all other components and will
  show up prominently in the weakest component which is the CO \cite{k2015}. However,
  even in CO, the SZ contamination is at the level of noise fluctuations in
  the rest of the map.
  Most of the foregrounds will thus be removed and suppressed even before we
  have applied the  ILC. The ILC
  should remove any remaining CO and dust foregrounds from the stacked
  maps bringing down the contamination to the noise levels. In
    particular  the ILC is most  efficient in removing
    the foregrounds in noiseless maps \cite[see][for
  proof and detailed study of ILC methods]{dc2007}. Stacking before ILC suppresses
    the noise thus increasing the efficiency of foreground removal.  We must still
account for the residual contamination and SZ signal from
  the stacked galaxies themselves. This is however not a big concern. Since the galaxies
  themselves 
  would be unresolved at Planck resolution, the dust and SZ signals from the
  galaxies is well approximated by two
non-overlapping Gaussian discs and subtracted.

With the parameter
fitting  exercise, we have therefore 
quantified the benefits of the {\it Stack First} approach. We have shown that
we have a factor of 1000 suppression in the fluctuating part of the
foregrounds, in addition to the suppression of noise,  by just stacking
the individual frequency maps even before the ILC component separation
algorithm is applied.

\section{Excess signal}\label{sec:excess_signal}

\begin{figure}
\includegraphics[width=1.05\linewidth]{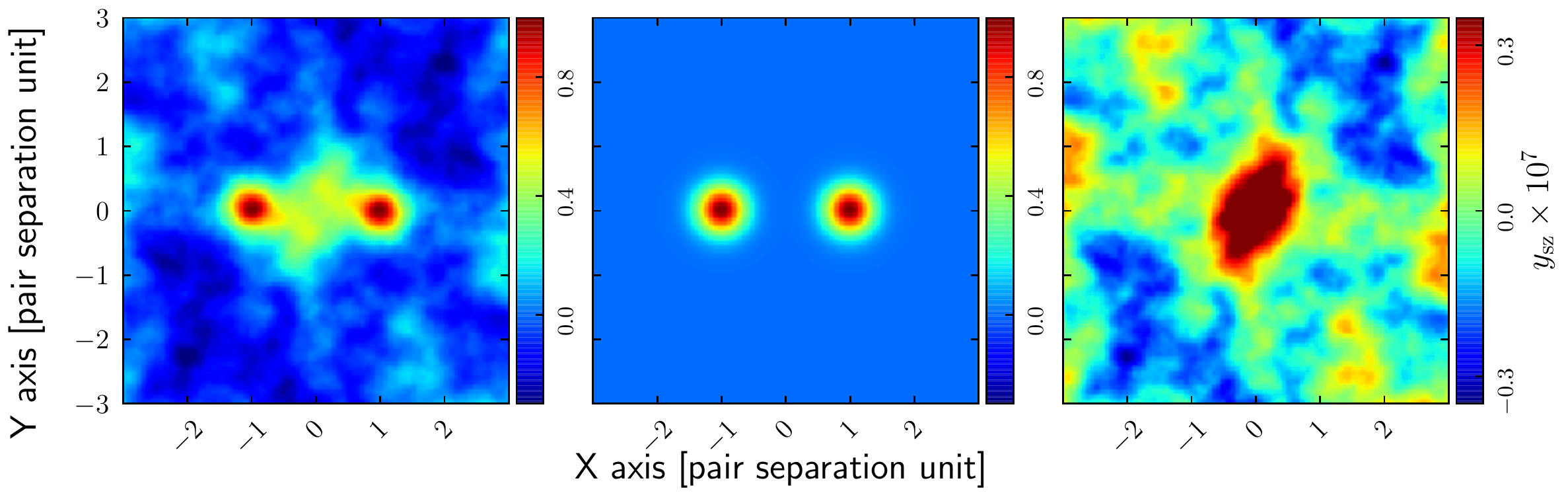}
\includegraphics[width=1.0\linewidth]{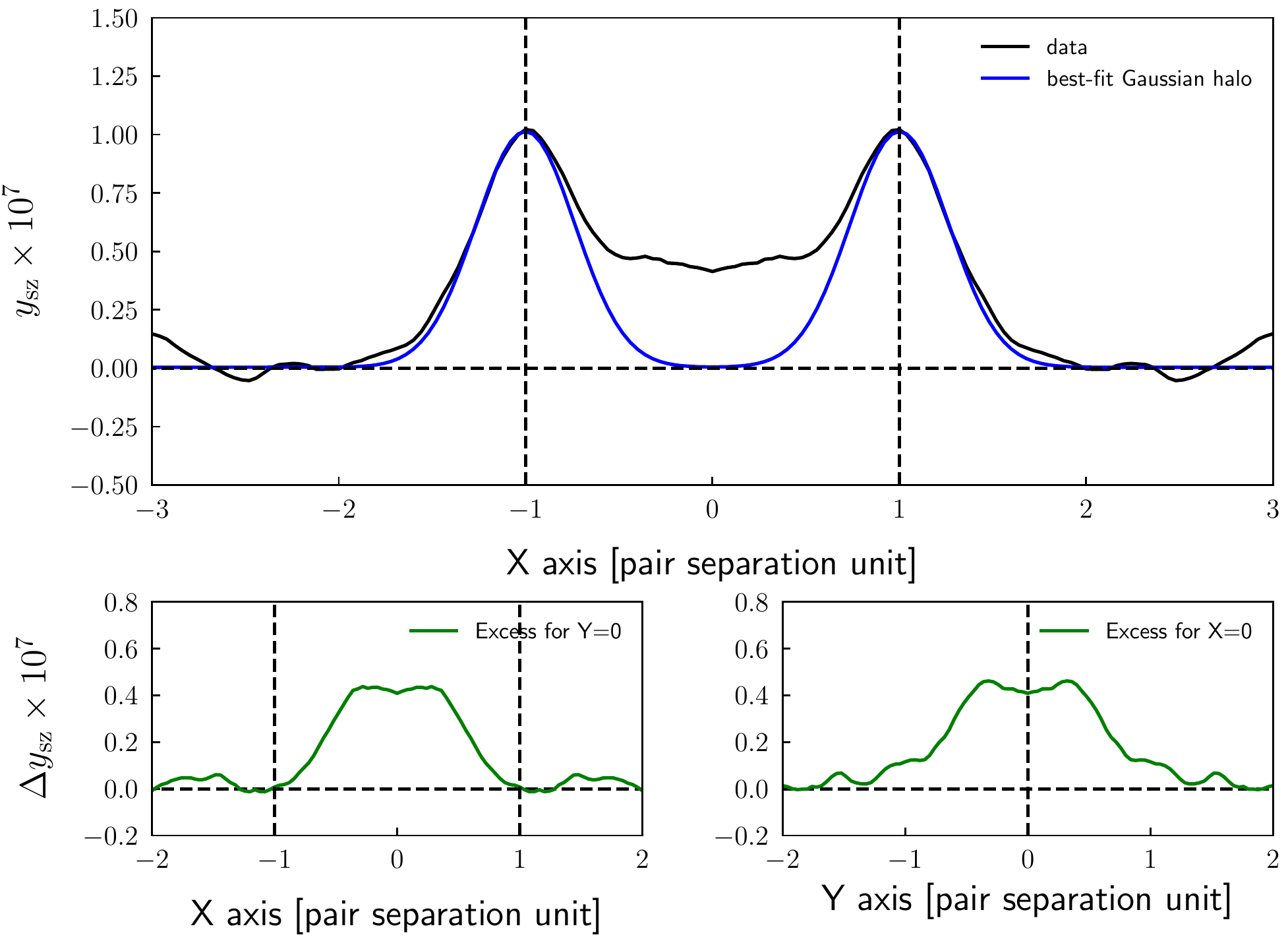}
\caption{\textit{Top Panel left:} The stacked ILC \hatysz\ map obtained from the
  combination of \planck\ LFI and HFI channel maps ($70-545$\,GHz) with the
  threshold criterion $\chisq \le 0.05$ and K86 mask. {\it Top Panel
    center:} The best-fit 2D gaussian model for the individual LRG halo
  contribution.  {\it Top Panel
    right:} The excess signal in the filament region connecting two LRGs
  after subtraction of the best-fit 2D Gaussian halo model from the stacked
  ILC \hatysz\ map. \textit{Middle panel} shows the \ysz profile  at Y=0,
  along with the best-fit Gaussian halo profile (blue solid
  line). \textit{Bottom panel left:} the excess at Y=0 after subtraction of
  best-fit model. \textit{Bottom panel right:}  the excess after
  subtraction of the best fit model at X=0.}
\label{fig:profile_stacked_ILC}
\end{figure}

We expect two symmetric peaks in the stacked ILC \hatysz\ map as we stack
the \planck\ frequency maps twice by interchanging the LRG positions.
The position of the LRGs may have some remaining dust contamination, since
 the LRGs themselves are expected to have strong dust emission (as evident by
the lower panels in the first column of Fig.~\ref{fig:stacked_point07_10}) and even a small  leakage
may be significant compared to the tSZ signal.
However, the region between the LRGs, where we expect to find the WHIM, is 
relatively free of dust and leakage from any weak dust emission from
the intergalactic medium would be suppressed even further by the ILC. The \textit{middle panel}
of Fig.~\ref{fig:profile_stacked_ILC} shows the profile of the stacked ILC
\hatysz\ map along Y=0 axis. The stacked ILC \hatysz\ signal for our
baseline case has the dominant contributions from the individual LRG
halo. The circular model for halos is a good approximation since most
  of our galaxies are unresolved and any non-circular beam effects will get symmetrized
  when stacking a large number of objects. \Tanimura\ have shown that other
  systematic effects are also small. To extract the excess \ysz\ signal in the filament region connecting the LRG pair, we need to subtract the individual LRG halo contribution.

We exclude the central region $-1<\text{X}< 1$ and fit a Gaussian model to the single-halo signal on both the sides \citep{Tanimura:2019}. The
blue solid line in the \textit{middle panel} of
Fig.~\ref{fig:profile_stacked_ILC} is the best-fit Gaussian model to the
data, excluding the central LRG region from $-1<\text{X}< 1$. The 2D
circular Gaussian model is constructed based on the fit to the data at
$Y=0$. The residuals after fitting the best-fit 2D circular Gaussian model
along the Y=0 and X=0 are shown in the \textit{bottom panel} of
Fig.~\ref{fig:profile_stacked_ILC}. The map representation of the same is
shown in the \textit{top panel} of Fig.~\ref{fig:profile_stacked_ILC}. The amplitude
of the excess \ysz\ signal is $\sim 4 \times 10^{-8}$ in the region $
-0.5 < \text{X} < 0.5$ at Y=0. The excess \ysz\ signal peaks in the central filament
region between $-0.5 < \text{X} < 0.5$ and $-0.5 <\text{Y}< 0.5$.  The average of
the excess \ysz\ signal in the central filament region is \ysz$=3.78\times10^{-8}$,
which is approximately  a factor of three higher than the value reported in T19.

We note that our basic selection criteria is same as \Tanimura\,
  however we impose additional thresholds to further prune our samples,
  making our results more robust to accidental contamination by foreground
  or background clusters.   In order to relate the measured \ysz\ signal to
  the filament properties, we use the following density profile for the filament
similarly to \Tanimura,
\begin{equation}
\Nez(r)=\frac{\Nez(0)}{\sqrt{1+\left(r / r_{\rm {c}}\right)^{2}}} \ ,    \label{eqn:5.2}
\end{equation}
where $\Nez(r)$ is the electron number density of a filament  at redshift
$z$ and distance $r$ from the filament center along the line of sight,  $r_c$ is the core radius of the
filament and we will take  $r_{\rm {c}}=0.5h^{-1}$\,Mpc. The
density profile is set to zero at $r > 5r_{\rm {c}}$.  The mean electron number
density as a function of redshift, $\bar{n}_{\rm e}(z)$ is given by 
$\bar{n}_{e}(z)=\frac{\Omega_{\rm b}\rho_{\rm cr} (1+z)^3}{\mu_{\mathrm{c}}
  m_{\mathrm{p}}}$, where $\Omega_{\rm b}$ is the baryon density parameter,
$\rho_{\rm cr}$ is the critical density of the Universe today, $m_{\rm p}$
is the proton mass and $\mu_{\rm
  c}=1.14$ is the mean molecular weight of primordial plasma with $76\%$
hydrogen by mass. We also define the overdensity $\delta$ at the center of
the filament as 
$ \delta=\Nez(0)/ \bar{n}_{\rm e}(z)$. We assume Planck $\Lambda$CDM
cosmology \cite{planck2018}  giving average electron number density today to
be 
$\barNe(0)=2.2\times 10^{-7}\,{\rm cm}^{-3}$.  
The electron number density in filaments, $\Nez(r)$  can be expressed in
terms of overdensity $\delta$ by multiplying and dividing Eq. \ref{eqn:5.2}
by $\bar{n}_{\rm e}(z)$ as
\begin{align}
\Nez(r)&= \frac{\Ne^z(0)}{\barNe(z)} \frac{\barNe(z)}{\sqrt{1+\left(r / r_{\rm {c}}\right)^{2}}} \nonumber \\ 
&=\delta \frac{\barNe(0)(1+z)^3}{\sqrt{1+\left(r / r_{\rm {c}}\right)^{2}}}. \label{eqn:5.3}
\end{align} 
Assuming a constant electron temperature $\Te$ and a symmetry along the filament axis, we can express $\Delta \ysz$ as a line of sight integration of the filament density profile ($\Ne(r,z)$) as,
\begin{align}
\Delta \ysz  &= \frac{\sigT \kB \Te}{\me c^{2}} \int^{5r_{\rm c}}_{-5r_{\rm c}}  \Nez(r) \id r \nonumber  \\ 
&= 2\frac{\sigT \kB \Te \delta (1+z)^3\barNe(0)}{\me c^{2}} \int^{5r_{\rm
    c}}_{0}    \frac{1}{\sqrt{1+\left(r / r_{\rm {c}}\right)^{2}}}   \id r
\ ,\nonumber\\
&=4.62r_{\rm c}\frac{\sigT \kB \Te \delta (1+z)^3\barNe(0)}{\me c^{2}}
\label{eqn:5.5}
\end{align}
where $5r_c$ is the cut-off radius of the filaments and $r$ is
the parameter for line of sight integration. The mean redshift  of
each LRG pair in our sample is known. On averaging over all LRG pairs in our fiducial
  sample, we obtain the average  excess \ysz\ signal from the filament for
  our sample as  \color{black}
\begin{equation}
\Delta \ysz= 3.78 \times 10^{-8} \left(\frac{\delta}{13}\right) \left(\frac{\Te}{5\times10^6}\right)\left(\frac{r_c}{0.5h^{-1}~ \text{Mpc}}\right).
\end{equation}
 
As the WHIM constitutes a major chunk of the matter in the filaments
\citep{Martizzi_Illustris:2019}, we can assume an average temperature of $5
\times 10^6$\,K in order to estimate the density contrast of the filament
from the \ywhim\ signal. This electron temperature  is within the upper
bounds obtained from IIlustris simulations \citep{Martizzi_Illustris:2019}
and in-between the temperatures used by \Tanimura\ ($\Te=10^7$\,K) and
\Graff\ ($\Te=10^6$\,K). Putting the $\ywhim=3.78 \times 10^{-8}$, we
obtain the mean overdensity in the filament region to be 
\begin{equation}
\delta  \approx 13 \left(\frac{5\times10^6~{\rm K}}{\Te}\right)\left(\frac{0.5 h^{-1} \text{Mpc}}{r_c}\right)
\end{equation}
Our result  is in excellent agreement with expectations
of overdensity in filaments ($\sim10-40$)  \citep{Cen_2001} from simulations.
We also use different electron density profiles to compute the mean over density
in the filament region. For electron density profile $\Ne(r) = {\rm constant} \ \ \  (r < 2r_c)$,
the mean overdensity is found to be $\sim 16$. For   $\Ne(r) = \frac{\Ne(0)}{1+\left(r / r_{\rm {c}}\right)^{2}} \ \ \ (r < 5
  r_c)$ density  profile, the mean overdensity is around 23. Irrespective of the electron density profile, the numbers for the overdensity only change by a factor of 2  and are always within the bounds of the expected WHIM overdensity as suggested by simulations.

\section{Estimate of error and significance of detection}\label{sec:null_tests}

 In order to obtain the significance of our detection we
 use the null test and the bootstrap method.

\subsection{Null test with misaligned stacking}\label{sec:error}

We use misaligned  LRG pairs, i.e. randomly chosen positions for the
  LRG pairs, to estimate the foreground contamination in the measured \ysz\
  signal. We make 100 random realizations of the  LRG pair catalogue. In
each realization, we shift the  Galactic longitude of every LRG pair in the
real data by a random amount $\in [5\deg ,25\deg]$ either in the positive
or negative direction, keeping the Galactic latitude fixed (for example, a
pair having central coordinated [$l,b$] could be shifted to
[$l+20\deg,b$]). The lower bound in the random shift in longitude makes
sure that the filament in the new random location does not overlap the
original filament and is sufficiently away from it. We keep the original
Galactic latitude so that the  Galactic foreground contamination is
  similar as the original location. If
indeed there was some contribution from the Galactic foregrounds, this
would be of the similar order of magnitude for the shifted pair, and hence
would show up as an excess in the misaligned stack too. We repeat the
procedure of component separation with ILC and measure the \ywhim\ in
each of the  100 random realizations of our
fiducial catalogue consisting of 88001 galaxy pairs. The results are shown
in Fig.~\ref{fig:misalignment_hist} \bs{(\textit{Left panel})}. We find the mean and standard deviation of the
WHIM  signal to be  $\ywhim= (-0.20 \pm 0.37)\times 10^{-8}$. We thus
find 
the detected \ywhim\ signal to be 
$10.2\sigma$ away from zero. Note that the systematic contributed by the mean of random realization, i.e.  the
background is $\approx 0.5 \sigma$ and is negative. We therefore neglect it
to get a conservative estimate of the detection significance.

\subsection{Bootstrap method}

We do an alternate estimate of the errorbar  on the reconstructed ILC \hatysz\
signal using the bootstrap technique. From our sample of $\Npairs=88001$,
we randomly select galaxy pairs to build a new sample, allowing each
galaxy pair to be sampled more than once, until we again have 88001 galaxy
pairs. Because of the
random selection, some galaxy pairs would be selected more than once while
some would be left out. We thus have a new realization of our galaxy pair
catalogue with the same number of pairs as in the original catalogue but  each galaxy pair, in general,  having a weight
different from unity.   We make 100 random realizations of our galaxy pair
catalogue in this way, and repeat our analysis by stacking, doing ILC component
separation and estimating the average \ywhim\ signal in the central region $
-0.5 < \text{X} < 0.5$ and $-0.5 <\text{Y}< 0.5$. The standard deviation
among the 100 realizations gives us an estimate of the sample variance or
the errorbar on the \ywhim\ signal. We find the  standard deviation for the
\ywhim\ signal in the region between the LRG pairs, i.e. $-0.5 < \text{X} < 0.5$ and $-0.5 <\text{Y}< 0.5$,  to be  $0.44\times
10^{-8}$, consistent with the misalignment method above.

\subsection{Non-overlapping misaligned stacking}
\begin{figure}
\centering
\begin{tabular}{cc}
\includegraphics[width=0.47\linewidth]{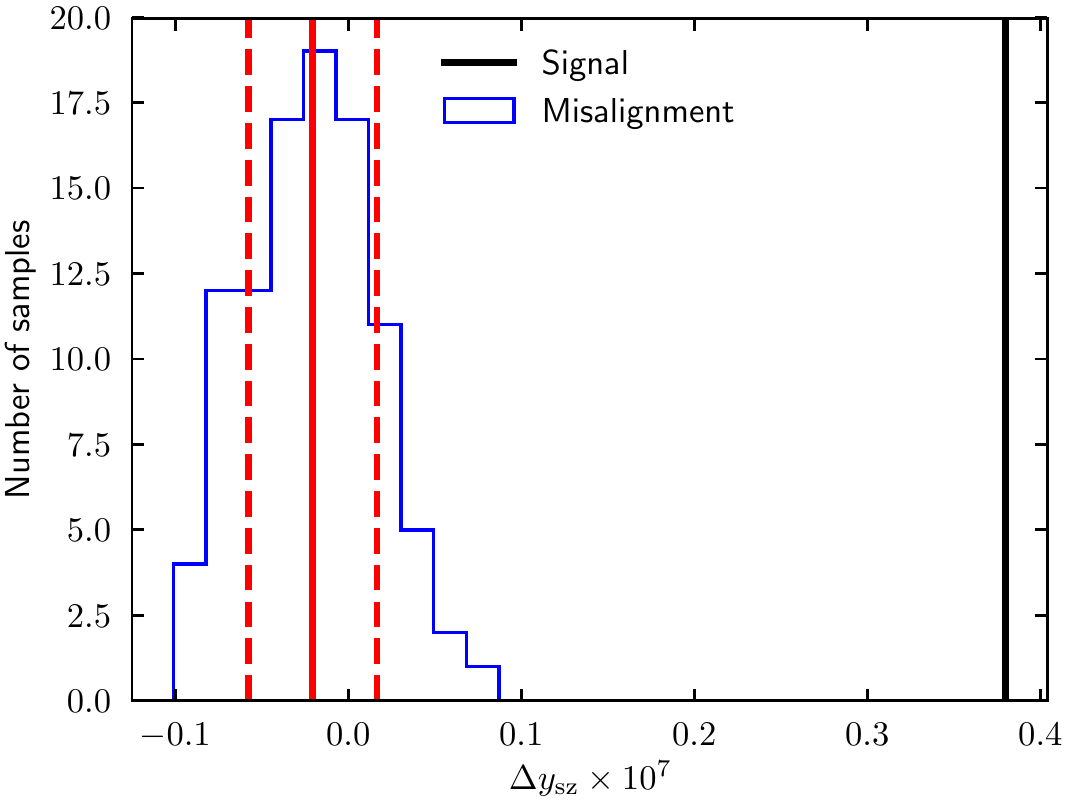} & \includegraphics[width=0.47\linewidth]{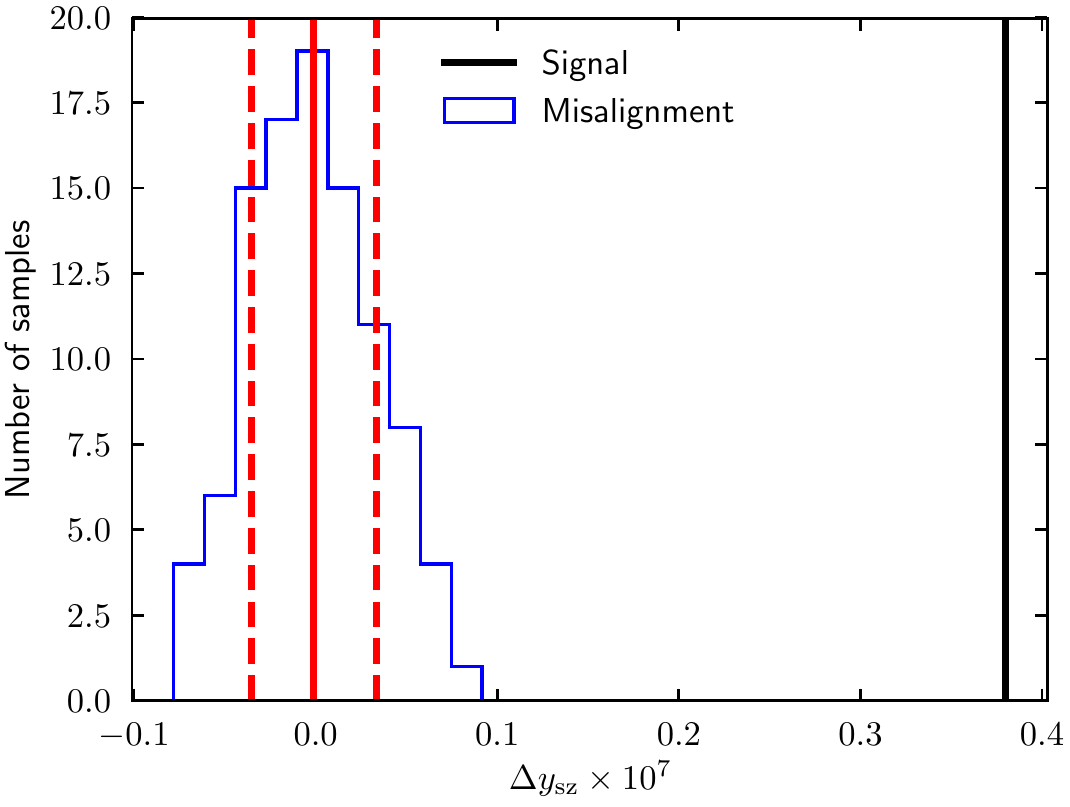}
\end{tabular}

\caption{\textit{Left panel:} The histogram of the excess \ysz\ signal obtained from random 100
  misaligned stacking is shown in blue colour and the black line represents
  the measurement from the actual data. Our measurement of $\ywhim=3.78\times
  10^{-8}$ is $10.2\sigma$ away from zero and $10.7\sigma$  away from the mean of the misaligned
  realizations. \textit{Right panel:} Same as the figure on left but with
  non-overlapping regions for each of the 100 realizations giving a
  detection significance of $10.8\sigma$.}
\label{fig:misalignment_hist}
\end{figure}

To address the concerns of the effect oversampling of regions might have on the estimate of our significance of results, we perform the misaligned stacking with mutually independent regions. For each pair in our sample concerned, with centre at [l,b], we choose 100 independent regions by dividing the iso-latitude region into 100 zones separated by ~3.5 degrees each. For each LRG pair, we assign values for misalignment such that there is no overlap for the 100 realizations in the entire region for $|X|<3$ and
$|Y|<3$. For example say the misalignment angle for two realizations for a pair are $k$ and $k^\prime$, so the patches centered at [l+$k$,b] and [l+$k^\prime$,b] never have any common region amongst them. The 100 realizations thus obtained for each of the 88000 misaligned LRG pairs
are mutually independent. The signal that we obtain from this analysis is
consistent within $1 \sigma$ with our previously quoted values where we did
not enforce the samples to be non-overlapping. The excess we obtain from
100 such misalignment samples is $\Delta y_{\rm sz}= (0.00 \pm 0.35) \times
10^{-8}$ and has been shown in Figure \ref{fig:misalignment_hist}
(\textit{Right panel}). With this our significance stands at $\sim 10.8
\sigma$  consistent with our other estimate in Sec. \ref{sec:error}. Since in trying to enforce non-overlap, we need to consider regions
which are far in longitude from the given galaxy pair, the contamination
may be slightly different in these samples giving a sligh difference in error
estimate compared to Sec. \ref{sec:error}.

\section{Consistency checks and robustness of the excess \ysz\ signal }

We perform a number of tests to check the robustness of our results and to
check that the excess \ysz\ signal that we observe is indeed coming from
the WHIM between the LRGs.

\subsection{Robustness w.r.t. resolution, channel combinations and selection
  criteria}

The WHIM signal is diluted to some extent due to the \planck\ beam
of 10\arcm. \Tanimura\ have shown with BAHAMAS simulations that there is
$\sim 15\%$ dilution in WHIM \ysz\ signal from unsmoothed maps to
10\arcm\ smoothed maps. As a consistency check, we  repeat our
analysis with \planck\ maps rebeamed to 8\arcm\ resolution. We would expect
the smaller beam size to confine the spread of LRG halo contribution to a
smaller region and thus reduce any contamination in the filament region. We
would also expect a slightly higher signal in the filament region between
the LRG pairs. We obtain the mean amplitude of the \ywhim\ signal in the
filament region to be $4.02\times10^{-8}$ at 8\arcm\ resolution. Thus the
dilution in the WHIM signal amplitude is $\sim  7\%$ due to beam
smoothing from 8\arcm\ to 10\arcm.

\begin{figure}
\centering
\includegraphics[width=1.0\linewidth]{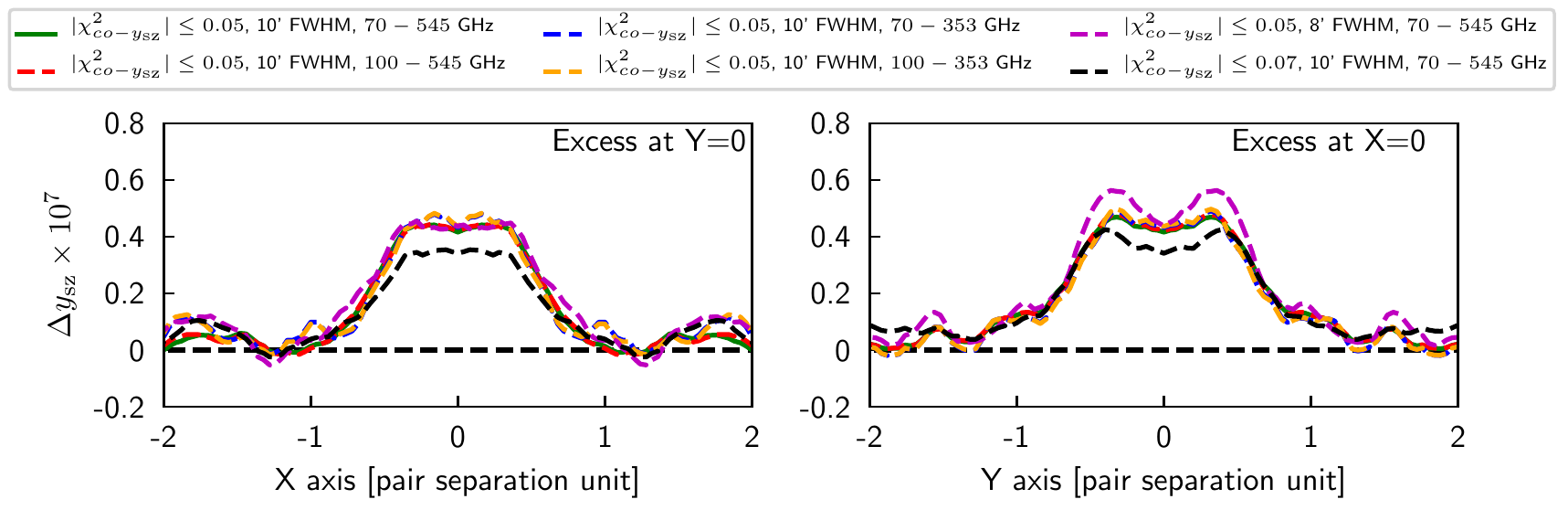}
\caption{The comparison of the excess \hatysz\ signals along X=0 and Y=0
  after the halo subtraction from individual LRG for different combinations
  of \planck\ channels  and  different beam resolutions.}
\label{fig:profile_comparison}
\end{figure}

We present the comparison of the WHIM profiles after halo subtraction for
  different selection of \planck\ channels, different \chisq\ thresholds,
  and resolutions  in Fig.~\ref{fig:profile_comparison}. The \ywhim\
  signal along Y=0 is consistent for different data selection
  criteria for $\chisq \leq 0.05$. As we see from Fig. \ref{fig:stacked_point07_10}, $\chisq \leq 0.07$ sample has significantly more contamination than the $\chisq \leq 0.05$ sample. This is also evident in the residual signal in Fig. \ref{fig:profile_comparison}, where we see that there is slightly smaller excess in-between the LRGs and slightly larger excess on the other sides of LRGs at $X,Y < -1$ and $X,Y > 1.$ \\

\subsection{Robustness w.r.t choice of point source masks and Galactic masks}

\begin{figure}
\centering
\includegraphics[width=1.0\linewidth]{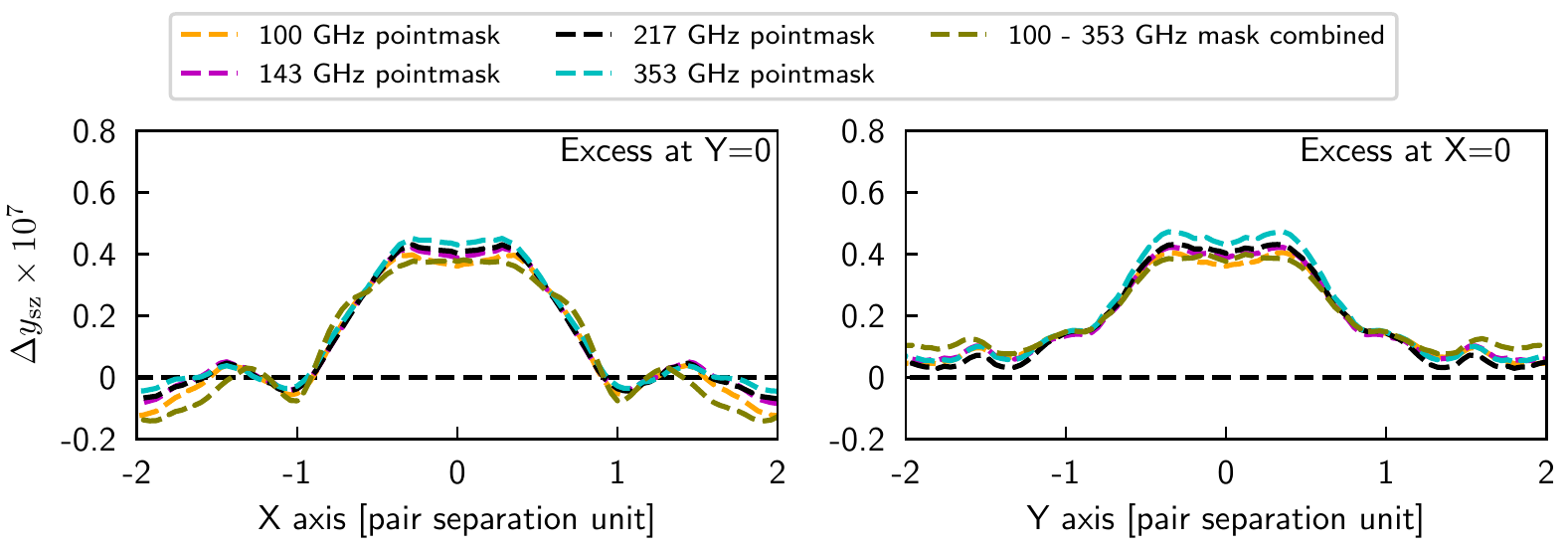}
\caption{Same as Fig.~\ref{fig:profile_comparison}, but for different
  combinations of \planck\ point source masks along with K86 mask with
  threshold criterion $|\chisq| \leq 0.05$ and  using $70-545$\,GHz \planck\ channels at 10\arcm\ beam resolution. }
\label{fig:profile_pointmask_comparison}
\end{figure}

The weights given to each frequency map during the ILC could vary significantly due to the presence of strong point sources in some of the regions projected for stacking analysis. Application of the temperature point source masks to these maps helps in removing those possible contaminants. We thus need to check the consistency of our excess by using point masks provided for the frequencies which get a significant weightage during ILC, i.e., the 4 HFI frequencies of 100-353 GHz most important for the \ysz\ extraction. We use the point source mask provided by  \planck\ to test the robustness
of the signal against contamination from strong radio and infrared point
sources. \planck\ provides individual temperature point source mask for
both LFI (30 - 70 GHz) and HFI (100 - 857 GHz)
\citep{Planck_A26_2016}. These are binary masks provided at \Nside=2048. We
downgrade them to \Nside\ = 1024 and then select the regions having values
$>$0.9 to ensure a sample free from point source contamination. These masks
are then combined with the K86 mask. We expect some variation in
\ywhim\ and $S/N$ as we are stacking a different number of LRGs
with different masks. If there is no contamination then the
\ywhim\ signal amplitude along Y=0 should remain within the sample
variance. Indeed, that is what we observe. We also test the robustness with the union of individual
frequency channel point masks combined with K86 mask as well. The results with different masks are presented in
Fig.~\ref{fig:profile_pointmask_comparison}. We do not see any significant
variation in the signal on using different masks.

\begin{figure}
\includegraphics[width=1.\linewidth]{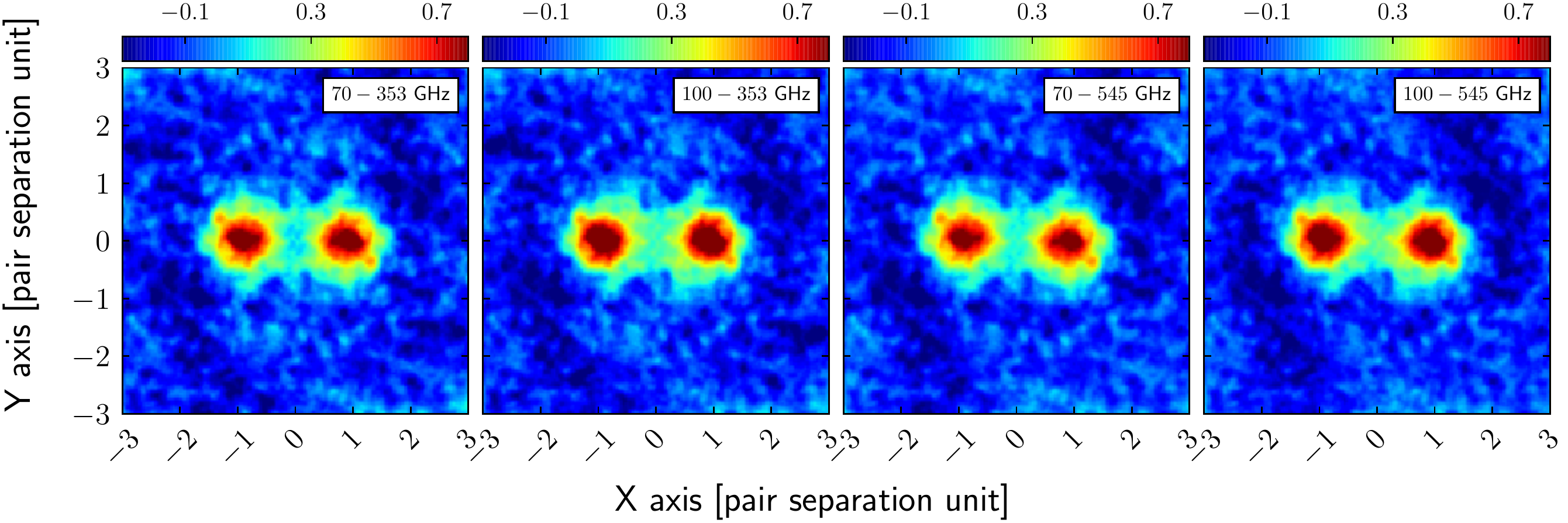}
\caption{The reconstructed ILC \hatysz\ maps using PL48 mask and different combination of \planck\ channels.} \label{fig:Stacked_ILC_PCM40}
\end{figure}

A point source mask has been provided by \planck\ \citep{Planck-A22:2016} for SZ
studies. However, we note that this is not intended as a general purpose mask but is
specifically customized
to be used with the NILC and MILCA SZ maps provided by the Planck
collaboration. We perform the stacking analysis of the \planck\ LFI and HFI channels
with PL48 mask which includes the point source mask. We stack roughly 101000 LRG
pairs retained by PL48. The number of LRG pairs allowed by the  PL48 mask is roughly the
same number as used in our baseline analysis with the threshold criterion
$\chisq \le 0.05$  with K86 mask. The reconstructed stacked ILC
\hatysz\ map using different combination of \planck\ channels is
presented in Fig.~\ref{fig:Stacked_ILC_PCM40}.  
The ILC weights for \planck\ frequency channels from
70 to 545 GHz for stacking with the PL48 mask are given in Table~\ref{tab:ilc_weights}. We
perform  all the same steps as we have done for the baseline analysis. The
average \ywhim\ signal in the 
central region $ -0.5 < \text{X} < 0.5$ and $-0.5 <\text{Y}< 0.5$ is
$(2.4\pm0.4)\times10^{-8}$. The error bar on the measured average \ywhim\ signal is derived from the
misalignment technique. 
If we also combine the PL48 mask with
K86 mask along with the threshold criterion $\chisq \le 0.05$ we retain
only $\approx 60000$ LRG pairs, i.e. only $\sim 60\%$ of the sample with
just PL48 mask. Thus the LRGs selected by our mask vs the PL48 mask are
 very different with PL48 mask sample giving a signal that is $\sim 36\%$
smaller. Our \ysz\ signal with PL48 mask sample is still much larger compared to the previous studies based on
stacking of \ysz\ maps and is in particular a $6\sigma$ detection.


\begin{figure}
\centering
\includegraphics[width=1.05\linewidth]{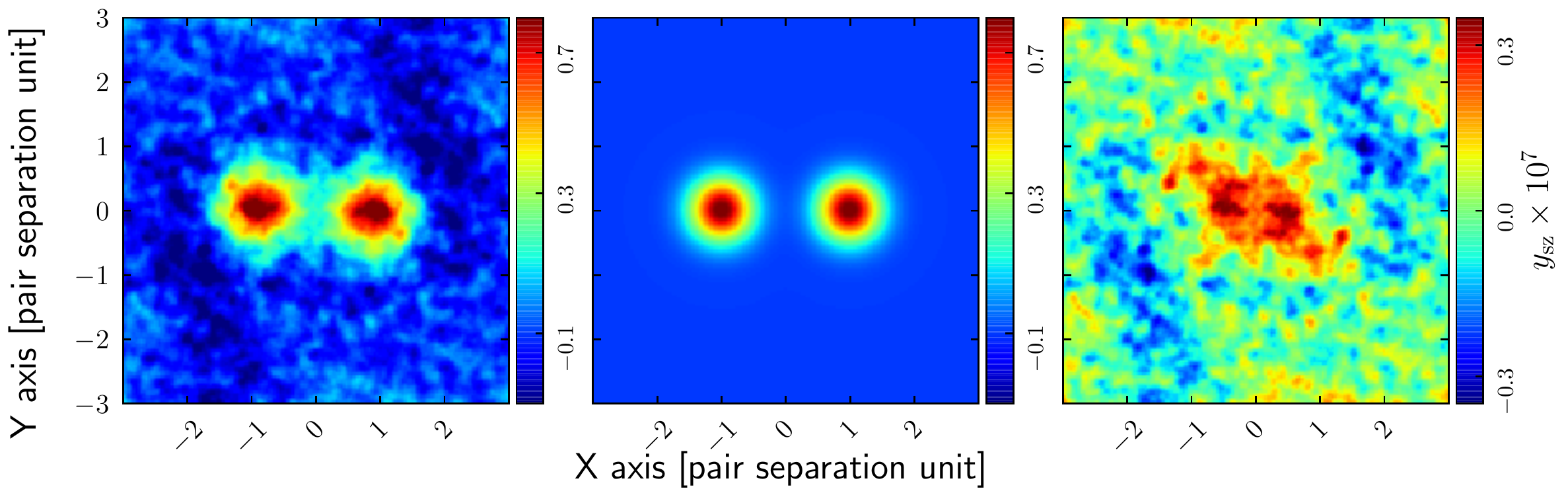}
\includegraphics[width=\linewidth]{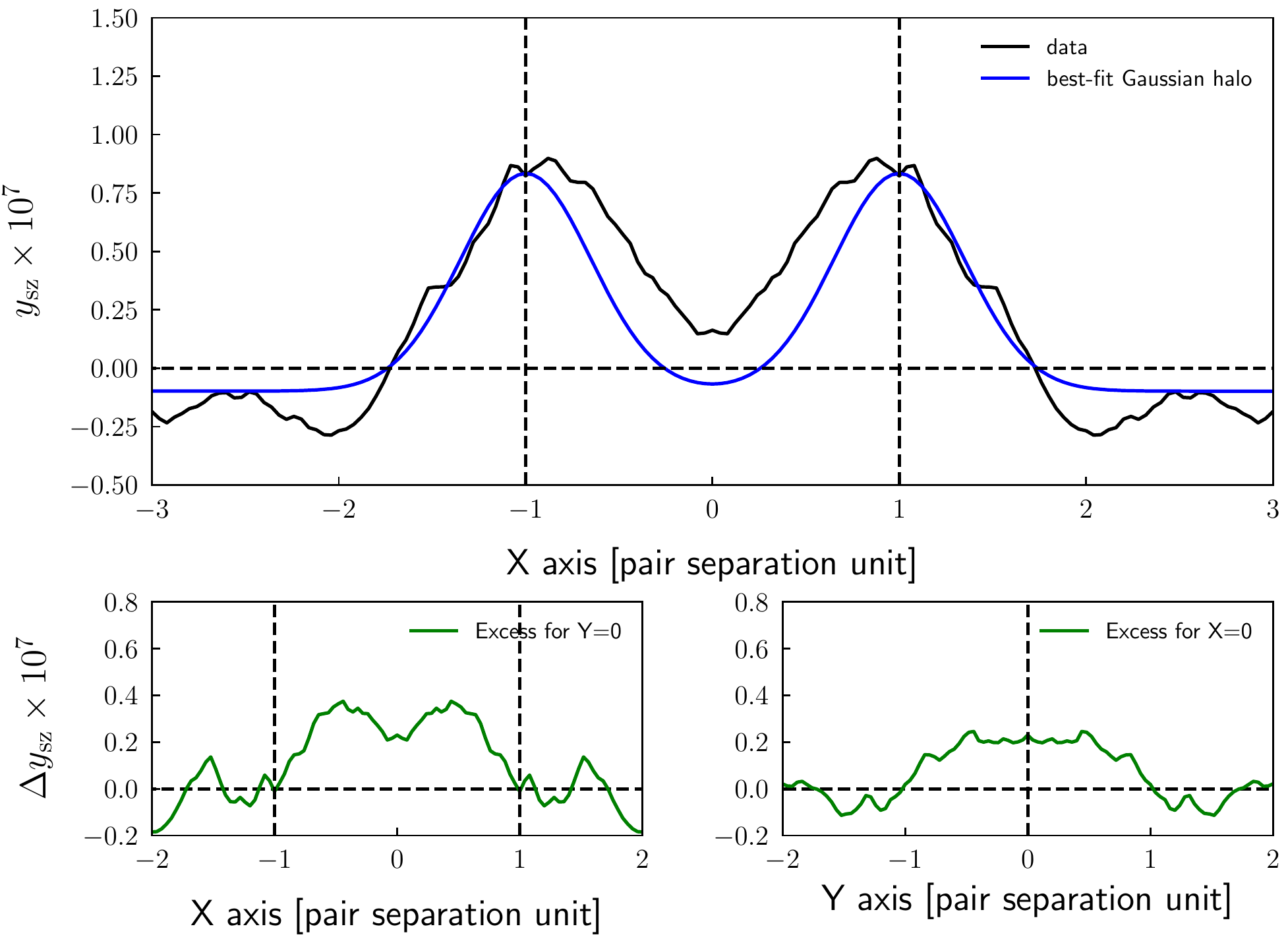}
\caption{Same as Fig.~\ref{fig:profile_stacked_ILC}, but for the stacked
  ILC \hatysz\ map using \planck\ HFI channels $70-545$\,GHz and PL48 mask.}
\label{fig:PCM40_profiles2}
\end{figure}

\begin{table}[h!]
\centering
\caption{The amplitude of the average \ywhim\ signal in the central filament region derived from different combinations of \planck\ channels in the reconstruct the ILC \hatysz\ map, different \chisq\ thresholds over K86 and for different FWHMs of the raw \planck\ maps. }
\label{tab:ilc_comparison}
\resizebox{0.8\textwidth}{!}{
\begin{tabular}{ |c|c|c|c|}
\hline
Channel combinations & \chisq threshold  & FWHM & \ywhim\  \\

 & with K86 & [in arcmin] & \\
\hline
$70-545$    & 0.05 & 10 & $3.78\times 10^{-8}$ \\
$100-545$  &  0.05 & 10& $3.78\times 10^{-8}$ \\
$70-353$  & 0.05 & 10  & $3.97\times 10^{-8}$ \\
$100-353$  & 0.05& 10 & $3.97\times 10^{-8}$\\
$70-545$ & 0.05 & 8  & $4.02\times 10^{-8}$\\
$70-545$ & 0.07& 10 & $3.50\times 10^{-8}$ \\
\hline
\end{tabular}
}
\end{table}

\section{Comparison with stacking of \planck\ \ysz\ maps \label{sec:comparision_yszmaps}}

Earlier \Tanimura\ and \Graff\ have stacked  the MILCA and NILC \ysz\ maps
 at the locations of the LRG pairs. As we argued earlier, the MILCA and NILC
 \ysz\ maps (or any other full sky tSZ map created from \planck\ data such as LIL
 map) have significant contamination (Fig.~\ref{fig:Comparision_diff_maps}), much higher compared to the
 signal we are interested in and there is no guarantee that the positive
 and negative contamination would cancel. As it turns out, there is
 over-cancellation, resulting in the excess negative contamination
 decreasing  the  \ysz\ signal in the WHIM between two LRGs.

\begin{figure}
\centering
\includegraphics[width=1.05\linewidth]{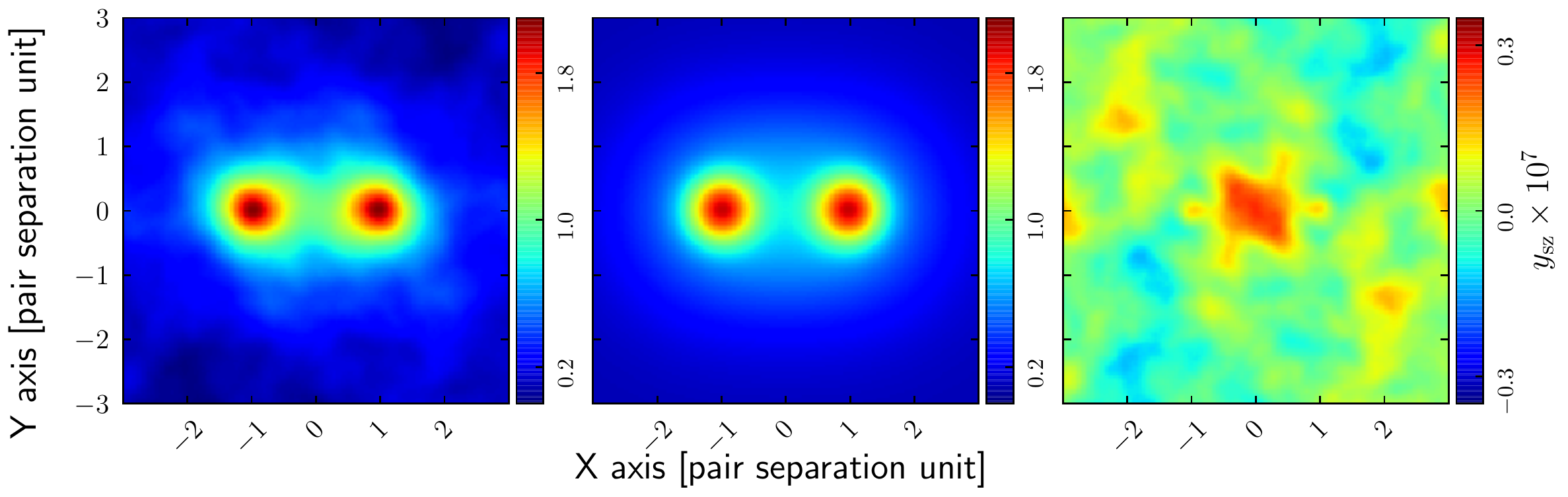}
\includegraphics[width=\linewidth]{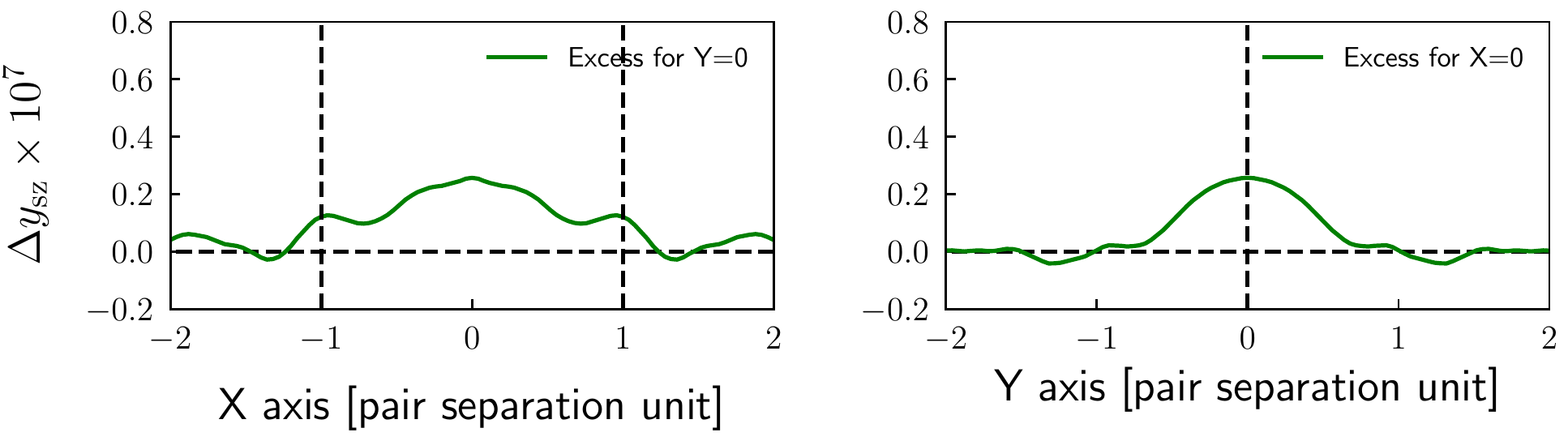}
\caption{Same as Fig.~\ref{fig:profile_stacked_ILC}, but for the stacked
  MILCA \ysz\ map with the threshold criterion $\chisq \le 0.05$ and K86 mask.}
\label{fig:MILCA_stack_profile}
\end{figure}

In this section we reproduce their results and in particular show that we
get results consistent with \Tanimura\ and \Graff\ when we also stack the
\ysz\ maps. We stack on the publicly available MILCA, NILC  and LIL \ysz\
maps at the location of LRG pairs  with K86 mask and selection threshold of
$|$\chisq$|< 0.05$. The stacking method  for \ysz\ maps is identical  to
the one we used for the  \planck\ frequency channel maps as described in
Sect.~\ref{sec:analysis}. We estimate the  local background signal for each
LRG pair  in the annular region $9 < r < 10$ ($r^2=X^2 + Y^2$) and subtract
it to get the excess \ysz\ signal above the background. We see
 two clear peaks at the position of LRG's
along with a bridge connecting the two peaks. The extended \ysz\ signal at
the peaks is due to the \ysz\ as well as galactic emission from the two
LRGs.  After subtracting the LRG halo contribution, we get average  $\ywhim \sim (1.88 \pm
0.30) \times 10^{-8}$ in the region between the LRGs, i.e. $-0.5 \leq \text{X} \leq 0.5$ and $-0.5 \leq
\text{Y} \leq 0.5$, from the stacked MILCA \ysz\  map.  The $1\sigma$ errorbar is
derived from the misalignment technique. The result of stacking analysis
from MILCA \ysz\ map is shown in Fig.~\ref{fig:MILCA_stack_profile}. 
We also use the other two publicly available \ysz\ maps for the stacking analysis.
Over the same filament region, we get the average $\ywhim \sim 1.56 \times 10^{-8}$ from NILC map
and $\ywhim \sim 3.45 \times 10^{-8}$ from LIL map. The WHIM signal we get from the stacking of LIL \ysz\ map
is close to the one we obtained from our ``{\it Stack First}" approach. 
 
\begin{figure}
\includegraphics[width=1.05\linewidth]{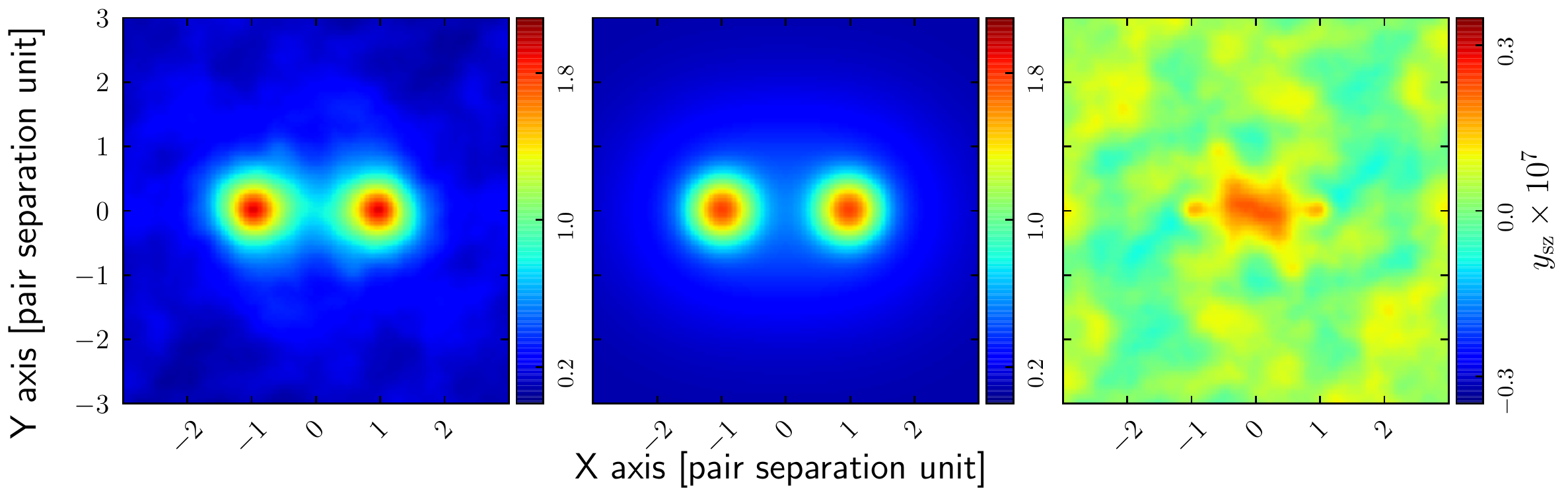}
\includegraphics[width=\linewidth]{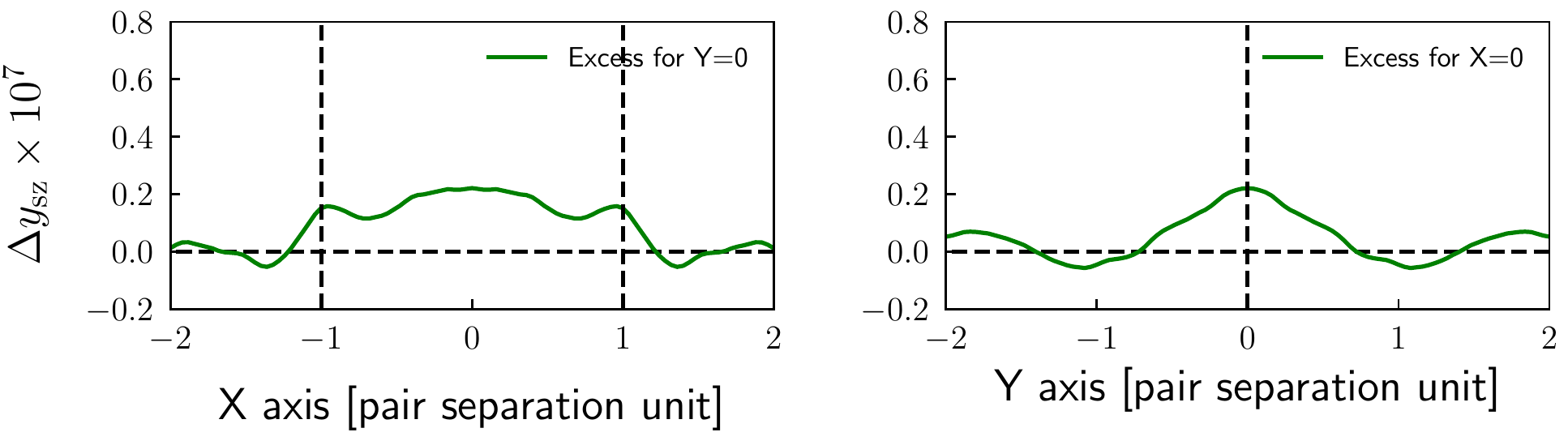}
\caption{Same as Fig.~\ref{fig:MILCA_stack_profile}, but for the stacked
  MILCA \ysz\ map and PL48 mask.}
\label{fig:Mask48_stack_MILCA}
\end{figure}

We also repeat the analysis of \Tanimura\ by using PL48. The results
  are presented in Fig.~\ref{fig:Mask48_stack_MILCA}. The WHIM \ysz\ signal
  in the filament region between the LRG pairs is $\ywhim= (1.55 \pm
  0.22)\times 10^{-8}$, consistent with the value we got above with our
  custom K86 mask and \chisq\ based selection criteria. Our results are also consistent within $1\sigma$ errorbars
  to the reported excess \ysz\ signal of $ (1.31\pm0.24)\times
  10^{-8}$ in T19. As argued above, and shown by our results from
  stacking the individual frequency maps, stacking \ysz\ maps gives biased
  results due to only partial cancelation  of the positive and negative
  contaminations.

\section{Conclusion}\label{sec:conclusion}

We have presented a new {\it Stack First} approach aimed at detection of weak \ysz\ signals in stacked objects in the \planck\ data. The important new
ingredient in our recipe is to {\it first stack} the individual frequency channel
maps and {\it then} perform blind component separation.  In our approach, the 
 component separation problem that the blind separation algorithm should
solve is simpler compared to the common method of first doing component
separation and then stacking. This is because the noise as well as the CMB
contribution is suppressed due to stacking at random locations and the stacked
dust signal becomes spatially uniform in its spectrum. In order to avoid
regions with CO contaminations,
present in all the \planck\ HFI channels excluding 143 GHz, as well as
strong background and foreground SZ sources, we use an
additional mask based on the \chisq\ thresholds. The \chisq\ map \citep{Khatri:2016} 
indicates whether CO emission or \ysz\ signal fits the \planck\ data
better in addition to the dust and CMB emission. For the weak, noise
dominated, sources we
are interested in, we should not be able to distinguish between CO and SZ,
and thus choose pixels with \chisq\ close to 0. We restrict our analysis to
a limited range of
\chisq\ values, i.e. $|\chisq| \leq 0.05$ to get a cleaner sample of LRG
pairs for the stacking analysis. Our approach is different than \Tanimura\
and \Graff, as they consider the stacking of the MILCA and NILC \ysz\ maps
at high Galactic latitude without taking into account the residual CO
emission and \ysz\ emission from the background/foreground sources in the \ysz\ maps. We present our main conclusions below. 

\begin{itemize}

\item We find the WHIM signal between the LRG pairs to be 
    $\ywhim= (3.78 \pm 0.37) \times 10^{-8}$. We have thus detected
  WHIM at a significance level of $\sim 10.2$ ($10.8\sigma$ if we use
  errorbar from non-overlapping regions).

\item We find the signal is robust with respect to using different channel
  combination and masks.

\item  Our results are consistent with the expectations
  for the WHIM from  hydrodynamic cosmological simulations
  \citep{Dave:2001, Cen_2001, Martizzi_Illustris:2019}. 
\item Our WHIM signal is higher compared to the results of 
  \Tanimura\ and \Graff. The difference is most likely coming from
  incomplete cancelation of positive and negative contamination when
  directly stacking \ysz\ maps.

\item The WHIM (dominant baryonic component in the filaments of the cosmic
  web), exists in a wide range of temperatures and densities. Assuming an
  average  temperature of $5\times10^6$\,K indicated by simulations, we
  find that the  overdensity in filaments is $\sim 13$ in agreement with
  the expectations from simulations  of $\sim 10-40$  \citep{Cen_2001}.
\end{itemize}

We therefore conclude that we have detected the missing baryons in the
local Universe using the thermal Sunyaev-Zeldovich effect in the \planck\ data at a
significance $>10\sigma$.

\section*{Acknowledgements}

BS thanks DST-INSPIRE and VSRP program of TIFR for the allowances which
made the visit to TIFR and stay for a part of the project possible. This
work was supported by Science and Engineering Research Board, Department of
Science and Technology, Govt. of India grant numbers
SERB/ECR/2018/000826 and ECR/2015/000078. This work was also supported by
Max Planck Partner group between Max Planck Institute for Astrophysics,
Garching and Tata Institute of Fundamental Research, Mumbai funded by
Max-Planck-Gesellschaft.  The computations in this paper were run on the
Aquila cluster at NISER supported by Department of Atomic Energy of the
Govt. of India. The \planck\ Legacy Archive (PLA) contains all public
products originating from the \planck\ mission, and we take the opportunity
to thank ESA/\planck\ and the \planck\ collaboration for the same. We thank
Prashant Bera for the useful discussion and help with the projection
code. Some of the results in this paper have been derived using the
\healpix\ package \cite{Gorski:2005}. We acknowledge support of the Department of Atomic Energy, Government of India, under project no. 12-R\&D-TFR-5.02-0200.

\bibliographystyle{unsrtads}
\bibliography{tsz_mdas.bib}

\appendix
\section{Validation of  {\it Stack First}  approach using simulations} \label{sec:appa}

In this section, we test the efficacy of the  {\it Stack First} approach
for the extraction of the \ywhim\ signal in the filament region connection
the LRG pairs. For this purpose, we use all-sky LIL \ysz\ map as a proxy
for the true tSZ signal and the extracted \ywhim\ signal from the stacking of LIL
\ysz\ map as our input or true WHIM signal for our fiducial galaxy pair
sample (K86 mask and $|\chisq| < 0.05$, see
Sect.~\ref{sec:comparision_yszmaps} for details). We prefer LIL \ysz\ map
over the MILCA and NILC \ysz\ maps just because of its agreement with the
amplitude of \ywhim\ signal obtained from the  {\it Stack First}
approach.

\begin{figure}
\centering
\includegraphics[width=0.75\linewidth]{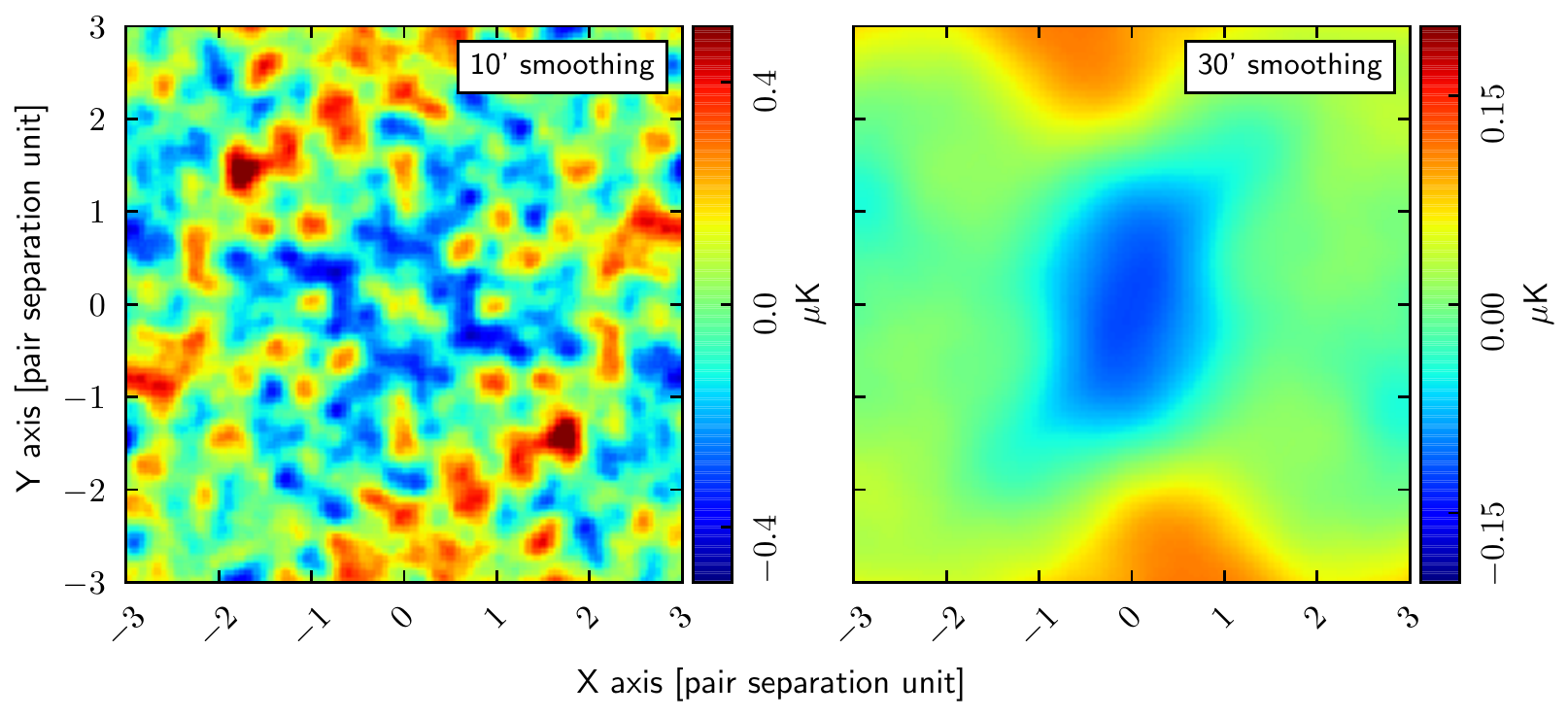}
\caption{The figure shows the stacked CO ($1\to 0$) map at 100 GHz for two different smoothing scales: 10\arcm\ FWHM (\textit{left panel}) and 30\arcm\ FWHM (\textit{right panel}) at the location of the LRG pairs for the threshold mask $|\chisq| <0.05$ with K86 mask. There is no significant correlation between the stacked CO ($1\to 0$) maps at two different beam resolutions and the stacked \ysz\ map. }
\label{fig:CO_stack1to0}
\end{figure}

We simulate realistic sky observed by \planck\ between 70 and 545 GHz by adding various diffuse foreground components to the input \ysz\ signal. We demonstrate that our {\it Stack First} approach recover back the input \ywhim\ signal from the combination of simulated multifrequency \planck\ maps without any significant bias. The total sky intensity at a given frequency $\nu$ can be modeled as a linear superposition of tSZ signal, CMB emission, the Galactic contamination and the instrumental noise,
\begin{align}
T (\nu, p)= f_{\nu}^{\ysz} T_{\ysz} (p) \,+ \, T_{\rm cmb} (p) \,+ \,  f_{\nu}^{\rm co} T_{\rm co} (p) \,  &+  f_{\nu}^{d} \tau_{353} (p) \left(\frac{\nu}{353}\right)^{\beta_{\rm d}(p)} B_{\nu} (T_{\rm d} (p))  \nonumber \\
&+ \,  T_{\rm sy} (\nu, p) \, + \, O_{\rm CIB} (\nu) \,  + \, T_{N} (\nu, p)  \ , \label{eqn:ModelSky}
\end{align}
where $p$ corresponds to the pixel index corresponding to \healpix\ pixel resolution of \Nside=1024. The different emission components that we included in our simulation are described below.
\begin{itemize}

\item tSZ signal - We use LIL \ysz\ map as a template for tSZ signal
  ($T_{\ysz}$). The factor $f_{\nu}^{\ysz}$ is the spectrum of the
  $y$-distortion. It is obtained by integrating the change in intensity
  ($\Delta I_{\nu}= \frac{2h\nu^3}{c^2}
  \frac{xe^x}{(e^x-1)^2}\left(\frac{x(e^x+1)}{(e^x-1)}-4\right) $) over the
  \planck\ frequency response \cite{planckhfi}, where $x=\frac{h\nu}{k_B T}$; $h,k_B, T$ being the Planck's constant, Boltzman constant and CMB monopole temperature (2.725 K) respectively. 

\item CMB emission - We simulate a random gaussian realization of the CMB
  sky from the theoretical power spectrum of the \planck\ 2018 best-fit
  model \cite{planck2018} giving the random  the CMB sky realization
  $T_{\rm cmb}$ uncorrelated with our input \ysz\ map.

\item Thermal dust - To simulate the dust emission, we use \planck\ 2013
  dust model \cite{Planck_A11_2014}, which is a modified blackbody fit to
  \planck\ intensity maps at $\nu \ge 353$\,GHz. The model fits the three
  parameters: dust opacity ($\tau_{353}$), dust spectral index ($\beta_{\rm
    d}$) and dust temperature ($T_{\rm d}$) per sky pixel. We use \planck\
  2013 dust model  \cite{Planck_A11_2014} instead of updated \planck\ 2015
  dust model \cite{Planck_A109_2016} since the 2013 model includes the CIB
  anisotropies present in the \planck\ maps. In the 2015 dust model, the
  CIB anisotropies are removed using the generalized linear combination
  (GNILC) method. As our study is focussed at high Galactic latitude and at
  small scales, we choose the appropriate dust model that is close to the
  real \planck\ data. Note that even though dust model in each pixel is
  quite simple, after stacking we will have superposition of many different
 spectra in each pixel. The foregrounds in the stacked maps, on which we
 run ILC, would be very complex and realistic.

\item Synchrotron - We use the bandpass integrated FFP10 synchrotron templates ($T_{\rm sy}$) available on PLA\footnote{\url{http://pla.esac.esa.int/pla}} for our purpose. It uses the 408 MHz map provided by Haslam \cite{Haslam:1982} and reprocessed by \cite{Remazeilles:2015} as a synchrotron amplitude and a pixel-dependent single power-law spectral index map, derived by fitting the \textit{WMAP} data \cite{Miville:2008}.

\item CIB offset - We add the CIB monopole term ($O_{\rm CIB}$) to our sky simulation. It has a constant value per frequency over the full-sky.  The CIB monopole values per \planck\ HFI frequency is given in Table 4 of \cite{Planck_L1_2018}.

\item CO emission - In absence of any publicly available CO map, we take
  the \planck-derived MILCA 2015 CO ($1 \to 0$) line emission map
  \cite{Planck_A1_2016} as a proxy for the CO emission. We choose Type 1 CO
  map as it has little contamination from other foregrounds, e.g. dust and
  \ysz\ emission. We smoothed the CO ($1 \to 0$) Type 1 map at 10\arcm\
  (FWHM) beam resolution taking into account the effective beam response of
  the \planck\ CO map and reduced to a \healpix\ resolution of
  \Nside=1024. We then stack the CO map at the location of LRG pairs with
  K86 mask and the \chisq\ threshold $|\chisq| < 0.05$. The left panel of
  Fig.~\ref{fig:CO_stack1to0} shows the stacked 10\arcm\ smoothed CO $J=1
  \to 0$ map at the location of LRG pairs. The stacked CO ($1 \to 0$) map
  has no structure similar to the \ysz\ signal and is completely dominated
  by the CO noise. This also confirms our expectation using parameter
  fitting in section \ref{sec:lil}.
  To increase the signal-to-noise ratio, we choose to smoothed the CO ($1 \to 0$) map to a beam resolution of 30\arcm\ FWHM. The stacking of 30\arcm\ smoothed CO ($1 \to 0$) map is presented in the right panel of Fig.~\ref{fig:CO_stack1to0}. The choice of 30\arcm\ smoothing scale is based on the power spectrum analysis of the CO map at high Galactic latitude \cite{Puglisi:2017}. We take 30\arcm\ smoothed CO ($1 \to 0$)  map as a template for the CO emission ($T_{\rm co}$). We add CO contribution to 100 and 217 GHz channels only. For 217 GHz channel, we assume the line ratio of $J=2 \to 1$ to $J=1 \to 0$ as 0.595. The factor $f_{\nu}^{\rm co}$ takes into account the spectrum of the CO emission at 100 and 217 GHz. For other frequencies, $f_{\nu}^{\rm co}$ is set to zero. 

\item Instrumental noise - We add a random realization of simulated full
  focal plane (FFP8) noise map per frequency as an instrumental noise
  contribution  $T_{N}$ \cite{Planck_A12_2016}. It captures the dominant instrumental, scanning and map making algorithm and implementation noise effects. 

\end{itemize}

\begin{figure}
\centering
\begin{tabular}{c}
\includegraphics[width=1.\linewidth]{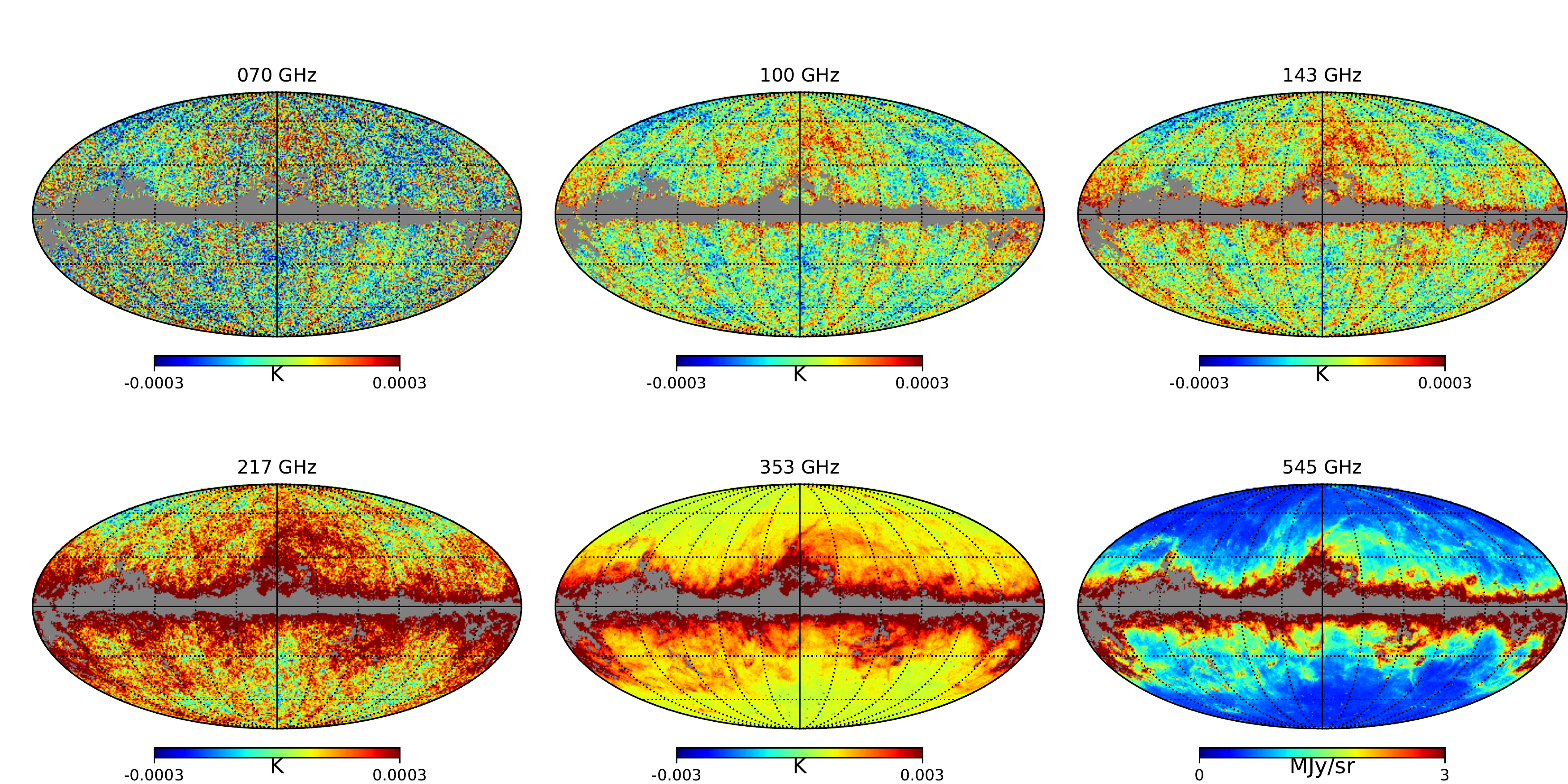}\\
(a)\\
\includegraphics[width=1.\linewidth]{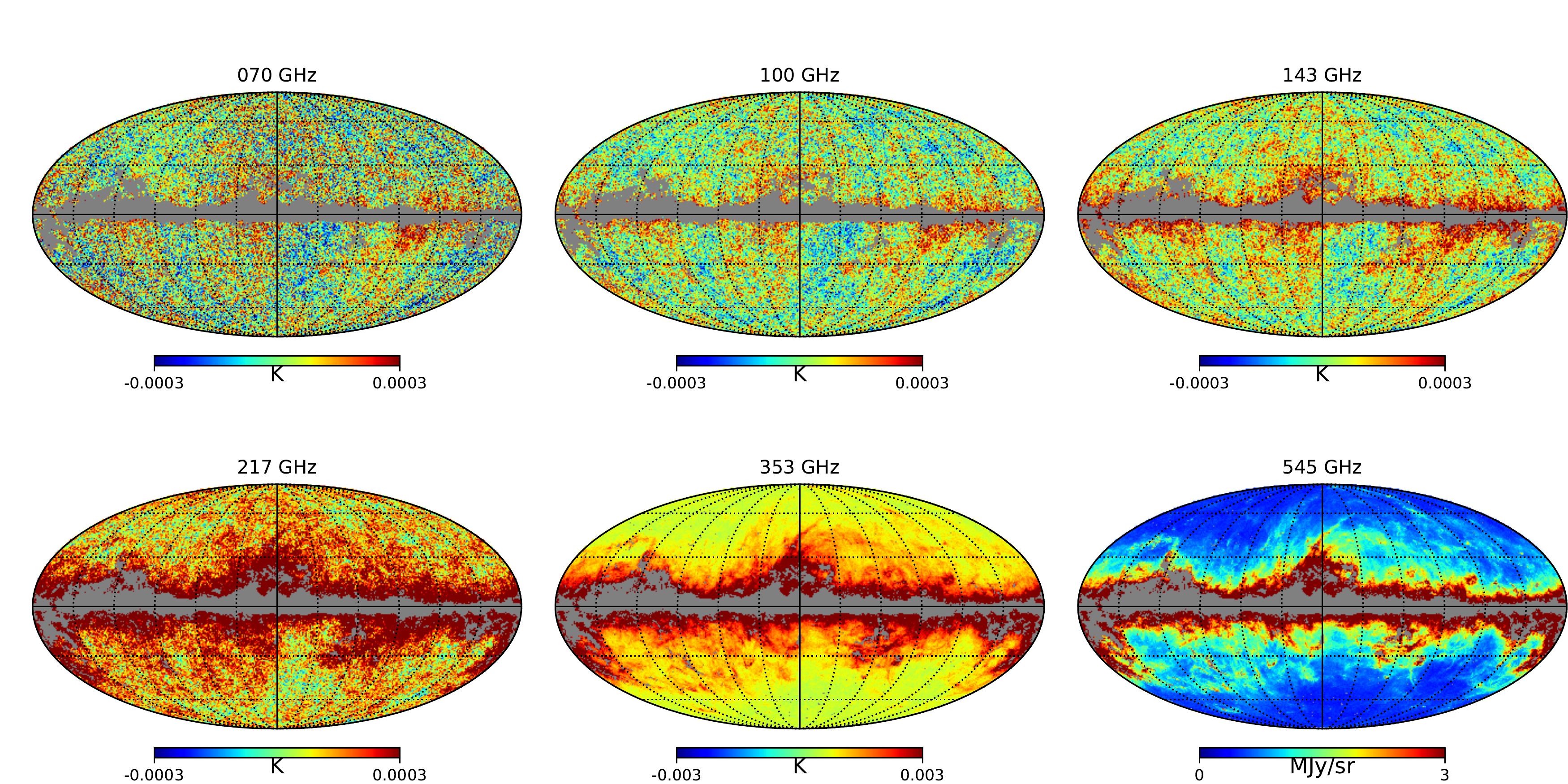}\\
(b)	
\end{tabular}
\caption{The figure shows the comparision of our simulated sky maps (\textit{panel a}) with the real \planck\ data (\textit{panel b}) in the frequency range between 70 and 545 GHz. All the maps are smoothed to 10\arcm\ FWHM beam resolution with K86 Galactic mask applied (shown in gray region).  \label{fig:plancksimscomp}}
\end{figure}

\begin{figure}
\includegraphics[width=\linewidth]{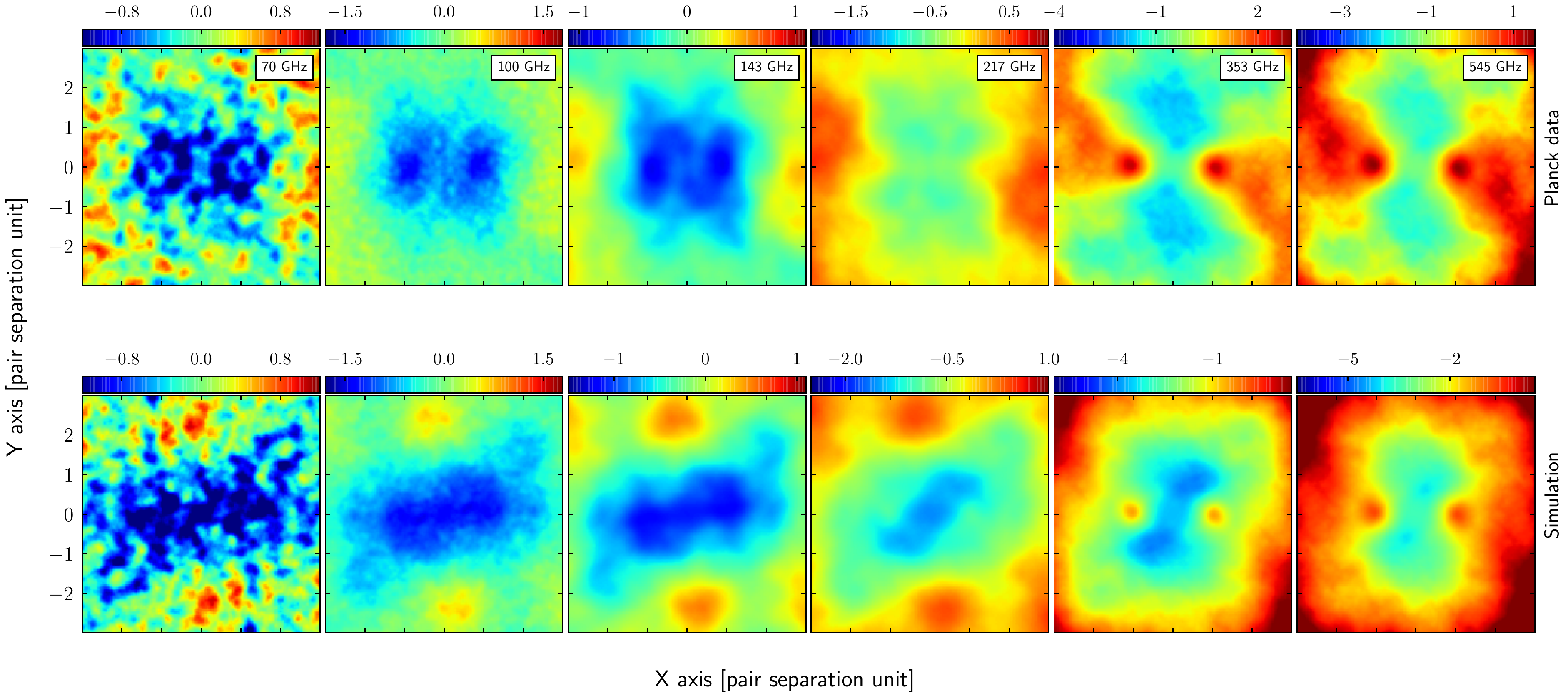}
\caption{The figure shows the comparison of stacked maps of the \planck\ data (\textit{top panel}) and our simulated dataset (\textit{bottom panel}) for \planck\ frequencies between 70 and 545 GHz. }
\label{fig:freq_simu}
\end{figure}

\begin{figure}
\includegraphics[width=\linewidth]{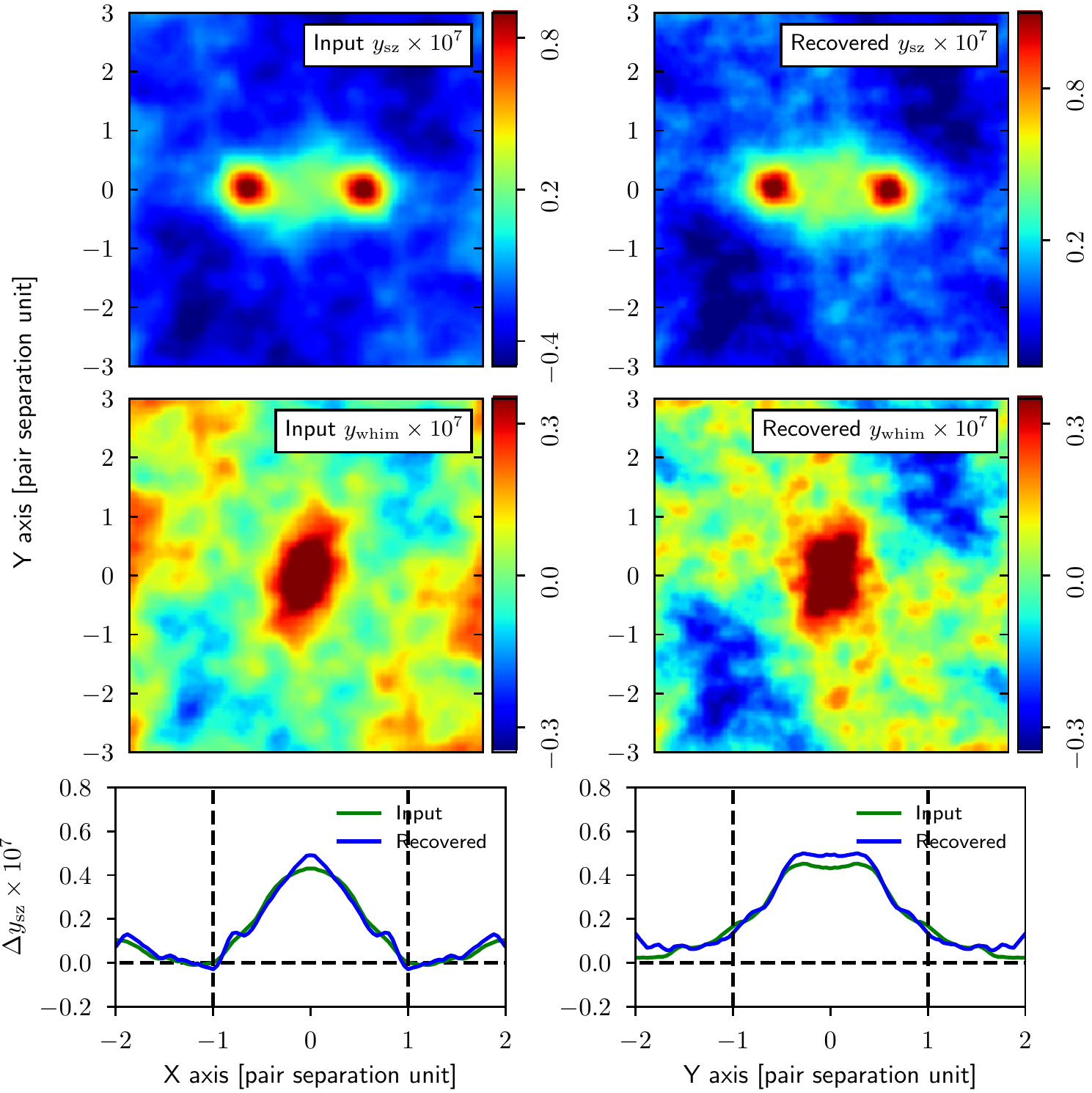}
\caption{The figure shows the comparison of the stacked input \ysz\ signal (\textit{top left panel}) and the recovered \ysz\ signal after applying the {\it Stack First}   approach (\textit{top right panel}) on the simulated \planck\ maps. The \textit{middle panel} shows the comparison between the input and the recovered \ywhim\ signal at the location of LRG pairs after subtraction of the halo contribution. The \textit{bottom panel} shows the profile of the input and recovered \ywhim\ signal along Y=0 (\textit{left panel}) and X=0 (\textit{right panel}). We are able to recover the profile of the \ywhim\ signal without any significant bias.}
\label{fig:residue_simulation}
\end{figure}

We simulate all-sky simulated maps between 70 and 545 GHz at a common beam resolution of 10\arcm\ FWHM. These maps are expressed in $K_{\rm CMB}$ units for frequencies between 70 and 353 GHz and \text{MJy/sr} for 545 GHz, similar to the real \planck\ data. Figure~\ref{fig:plancksimscomp} compares the simulated maps over K86 mask with the real \planck\ data. We then perform the {\it Stack First}  approach i.e. stacking the simulated frequency maps first at the LRG pair location over the threshold mask ($|\chisq| < 0.05$) with K86 mask and then perform the ILC on the stacked maps. Figure~\ref{fig:freq_simu} shows the comparison between the simulated and the \planck\ stacked maps at frequencies between 70 and 545 GHz. The results and comparision with the stacked input map has been shown in the Figure \ref{fig:residue_simulation}. We find that the \ywhim\ signal recovered using the {\it Stack First} approach indeed resembles the input \ywhim\ signal obtained from the direct stacking of the LIL \ysz\ map. This confirms beyond doubt the capabilities of our approach in recovering faint signals by removing contaminations from the stacked frequency maps.

\end{document}